\documentclass[aps,prd,twocolumn,preprintnumbers,
superscriptaddress,nofootinbib]{revtex4-2}
\usepackage{dcolumn}
\usepackage{amsmath}
\usepackage{amssymb}
\usepackage{amsthm}
\usepackage{mathtools}
\usepackage{mathrsfs}
\usepackage{bm}
\usepackage{slashed} 
\usepackage{graphicx}
\usepackage{multirow}
\usepackage{tikz}

\usepackage{natbib}
\usepackage[caption=false]{subfig}
\usepackage{relsize}	
\usepackage{array}
\usepackage{float}
\usepackage{color}
\usepackage{xcolor}
\usepackage{soul}
\usepackage{verbatim} 
\usepackage{siunitx}

\usepackage{hyperref}
\usepackage{CJKfntef}
\usepackage{ulem}
\hypersetup{colorlinks=true,linkcolor=purple,anchorcolor=blue,citecolor=blue, filecolor=blue,urlcolor=red,bookmarksnumbered=true,
	pdfview=FitB
}
\allowdisplaybreaks
\def\be{\begin{eqnarray}}
	\def\ee{\end{eqnarray}}

\colorlet{purple1}{blue!70!red}
\colorlet{darkred}{red!50!black}

\begin{document}
	
	\title{Double parton distributions of the proton from basis light-front quantization}
	
	\author{Tian-Cai Peng}
	\email{pengtc20@lzu.edu.cn} 
	\affiliation{School of Physical Science and Technology, Lanzhou University, Lanzhou 730000, China}
	\affiliation{Lanzhou Center for Theoretical Physics, Key Laboratory of Theoretical Physics of Gansu Province, Lanzhou University, Lanzhou 730000, China}
	\affiliation{Key Laboratory of Quantum Theory and Applications of MoE, Lanzhou University, Lanzhou 730000, China}
	\affiliation{Research Center for Hadron and CSR Physics, Lanzhou University and Institute of Modern Physics of CAS, Lanzhou 730000, China}
	
	\author{Zhi~Hu}
	\email{huzhi0826@gmail.com} 
	\thanks{*Corresponding author}
	\affiliation{High Energy Accelerator Research Organization (KEK), Ibaraki 305-0801, Japan}

	\author{Sreeraj~Nair}
	\email{sreeraj@impcas.ac.cn} 
	\affiliation{Institute of Modern Physics, Chinese Academy of Sciences, Lanzhou 730000, China}
	\affiliation{School of Nuclear Science and Technology, University of Chinese Academy of Sciences, Beijing 100049, China}
	\affiliation{CAS Key Laboratory of High Precision Nuclear Spectroscopy, Institute of Modern Physics, Chinese Academy of Sciences, Lanzhou 730000, China}
	
	\author{Siqi~Xu}
	\email{xsq234@impcas.ac.cn} 
	\affiliation{Institute of Modern Physics, Chinese Academy of Sciences, Lanzhou 730000, China}
	\affiliation{School of Nuclear Science and Technology, University of Chinese Academy of Sciences, Beijing 100049, China}
	\affiliation{CAS Key Laboratory of High Precision Nuclear Spectroscopy, Institute of Modern Physics, Chinese Academy of Sciences, Lanzhou 730000, China}

	\author{Xiang~Liu}
	\email{xiangliu@lzu.edu.cn}
	\affiliation{School of Physical Science and Technology, Lanzhou University, Lanzhou 730000, China}
	\affiliation{Lanzhou Center for Theoretical Physics, Key Laboratory of Theoretical Physics of Gansu Province, Lanzhou University, Lanzhou 730000, China}
	\affiliation{Key Laboratory of Quantum Theory and Applications of MoE, Lanzhou University, Lanzhou 730000, China}
	\affiliation{Research Center for Hadron and CSR Physics, Lanzhou University and Institute of Modern Physics of CAS, Lanzhou 730000, China}
	\affiliation{MoE Frontiers Science Center for Rare Isotopes, Lanzhou University, Lanzhou 730000, China}

	\author{Chandan~Mondal}
	\email{mondal@impcas.ac.cn} 
	\affiliation{Institute of Modern Physics, Chinese Academy of Sciences, Lanzhou 730000, China}
	\affiliation{School of Nuclear Science and Technology, University of Chinese Academy of Sciences, Beijing 100049, China}
	\affiliation{CAS Key Laboratory of High Precision Nuclear Spectroscopy, Institute of Modern Physics, Chinese Academy of Sciences, Lanzhou 730000, China}
	
	\author{Xingbo~Zhao}
	\email{xbzhao@impcas.ac.cn} 
	\affiliation{Institute of Modern Physics, Chinese Academy of Sciences, Lanzhou 730000, China}
	\affiliation{School of Nuclear Science and Technology, University of Chinese Academy of Sciences, Beijing 100049, China}
	\affiliation{CAS Key Laboratory of High Precision Nuclear Spectroscopy, Institute of Modern Physics, Chinese Academy of Sciences, Lanzhou 730000, China}
	\affiliation{Advanced Energy Science and Technology Guangdong Laboratory, Huizhou, Guangdong 516000, China}
	
	\author{James~P.~Vary}
	\email{jvary@iastate.edu}
	\affiliation{Department of Physics and Astronomy, Iowa State University, Ames, IA 50011, U.S.A.}

	\collaboration{BLFQ Collaboration}
	
	\date{\today}
	
	\begin{abstract}
		Within the basis light-front quantization framework, we systematically investigate the unpolarized and longitudinally polarized double parton distributions (DPDs) of quarks inside the proton. We utilize the light-front wave functions of the proton derived in the valence sector from a Hamiltonian quantized on the light-front. The interaction terms of the Hamiltonian consist of a one-gluon exchange interaction at fixed coupling and a three-dimensional confinement potential. Our current analysis yields significant correlations of the quarks' longitudinal momenta with their transverse separation.  We also demonstrate that our calculations do not support the commonly used $x-\vec{k}_\perp$ factorization of the DPDs in $x$ and $k_\perp$.
		Our results are qualitatively consistent with those of other phenomenological models.
		
	\end{abstract}
	
	\maketitle
	\section{Introduction} \label{intro}
	The parton distribution functions (PDFs) are essential for describing the nonperturbative structure of hadrons in high-energy collider measurements, such as those conducted at the Large Hadron Collider (LHC).
	The PDFs give the probability of finding a parton carrying a certain momentum fraction $x$ inside the hadron probed by single parton scattering~\cite{Soper:1996sn}. 
	In general, within the same hadron-hadron collision, there is the possibility for multiple hard scattering events and in fact, the contributions from such multiparton interactions (MPIs)~\cite{Bartalini:2018qje} increases with increasing energy. The most important and the simplest of these MPIs is double parton scattering (DPS), wherein two partons in each hadron participate in two distinct hard scattering processes. These DPS cross-sections can be parametrized in terms of the nonperturbative distributions called the double parton distributions (DPDs), which effectively quantify the joint probability of finding the two struck partons with longitudinal momentum fraction $x_1$ and $x_2$, and separated by a transverse distance $\vec{y}_\perp$ inside the hadron. Most DPS studies assume that there is no correlation between the transverse separation and the longitudinal momentum fractions~\cite{CMS:2022pio, Leontsinis:2022cyi, CMS:2021lxi, LHCb:2020jse, CMS:2019jcb, Belyaev:2017sws, CMS:2017han, Belyaev:2017sws, An:2017kyn}. A complete factorization of all the arguments leads to the so-called “pocket formula”~\cite{Koshkarev:2022mgi}, wherein the DPS cross section is written as the product of two single parton cross-sections. 
	
	The validity of the factorization hypothesis will be examined here through analysis of di-quark correlations in a light-front Hamiltonian framework. We note  that the DPDs depend not only on the longitudinal momentum fraction and the transverse separation but they also depend on the renormalization scales $\mu_1$ and $\mu_2$ corresponding to the two struck partons. They obey the QCD evolution equation~\cite{Kirschner:1979im,SHELEST1982325,Zinovev:1982be,Snigirev:2003cq,Korotkikh:2004bz,Ceccopieri:2010kg,Ceccopieri:2014ufa} similar to the 
	Dokshitzer-Gribov-Lipatov-Altarelli-Parisi (DGLAP) evolution equations for the PDFs. The evolution of DPDs generally neglects the color correlation~\cite{Mekhfi:1985dv} between the two partons based on the fact that they are Sudakov suppressed~\cite{Mekhfi:1988kj,Manohar:2012jr} and instead, the color of each individual parton is summed over. Meanwhile, the color-correlated evolution kernel of the DPDs at the next-to-leading order in the strong coupling has been computed in Ref. \cite{Diehl:2022rxb}.
	
	Experimentally, DPS events have been observed at the Tevatron and the LHC for a wide range of final states~\cite{CDF:1997yfa, D0:2015rpo, LHCb:2016wuo, ATLAS:2019jzd, CMS:2019jcb, CMS:2021lxi} with future LHC data expected to yield better insights. In upcoming electron-ion collision experiments~\cite{AbdulKhalek:2021gbh, Anderle:2021wcy}, it may also be feasible to measure DPDs by examining the interaction between two photons and the nucleon~\cite{Katich:2013atq, Metz:2012ui, Afanasev:2007ii}.
	
	The foundational theory on DPS was laid several decades ago providing remarkable insights ~\cite{Paver:1982yp, PhysRevD.32.2371, SJOSTRAND1987149}. Several studies have been performed in the last decade to improve our theoretical understanding of DPS
	~\cite{Blok:2010ge, Gaunt:2011xd, Ryskin:2011kk,Blok:2011bu, Diehl:2011yj, Manohar:2012jr, Manohar:2012pe,Ryskin:2012qx}. The DPDs are constrained by sum rules based on quark number and conservation of momentum~\cite{Gaunt:2009re, Golec-Biernat:2014bva, Golec-Biernat:2015aza, Diehl:2020xyg} and also from their nature in the limit of small distances between the two struck partons~\cite{Diehl:2011tt, Diehl:2019rdh}. Several studies exist on DPDs in QCD-inspired models~\cite{Chang:2012nw,Rinaldi:2013vpa,Broniowski:2013xba,Rinaldi:2014ddl,Broniowski:2016trx,Kasemets:2016nio,Rinaldi:2016jvu,Rinaldi:2016mlk,Rinaldi:2018zng,Courtoy:2019cxq,Broniowski:2019rmu,Rinaldi:2020ybv,Courtoy:2020tkd,Mondal:2019rhs}. The DPS is also considered for physics beyond the Standard Model~\cite{CMS:2020cpy} as well as in the background of precision studies~\cite{DelFabbro:1999tf}. A few of the recent investigations both from theory and phenomenology can be found in Refs.~\cite{Fedkevych:2020cmd, Ceccopieri:2021luf,Blok:2022mtv, Golec-Biernat:2022wkx,Diehl:2023jje, Diehl:2022dia, Diehl:2023cth}. Additionally, Euclidean lattice QCD stands out as a promising theoretical framework for calculating DPDs of the nucleon~\cite{Bali:2021gel, Zimmermann:2022emx, Zhang:2023wea}.
	
	In this work, we adopt the basis light-front quantization (BLFQ) approach~\cite{Vary:2009gt}, a recently developed nonperturbative framework, to calculate the DPDs of the proton. With the framework of BLFQ, the light-front wave functions (LFWFs) of a bound state are obtained by diagonalizing the light-front Hamiltonian in a truncated Fock space~\cite{Brodsky:1997de}. The BLFQ approach has been successfully employed in QED systems~\cite{Honkanen:2010rc, Zhao:2014xaa, Wiecki:2014ola, Chakrabarti:2014cwa, Hu:2020arv, Nair:2022evk, Nair:2023lir} as well as in several QCD bound systems~\cite{Jia:2018ary, Lan:2019vui, Lan:2019rba, Adhikari:2021jrh, Lan:2021wok, Mondal:2021czk, Li:2015zda, Li:2017mlw, Li:2018uif, Lan:2019img, Tang:2018myz, Tang:2019gvn, Mondal:2019jdg,Xu:2021wwj, Liu:2022fvl, Hu:2022ctr, Peng:2022lte,Kaur:2023lun, Zhu:2023nhl,Zhang:2023xfe, Liu:2024umn,Kaur:2024iwn, Yu:2024mxo}. Recent studies within BLFQ that go beyond the valence Fock sector, thereby incorporating gluon dynamics, can be found in Refs.~\cite{Kaur:2024iwn,Xu:2023nqv, Lin:2023ezw,Zhu:2023lst, Yu:2024mxo}. Here, we adopt a light-front effective Hamiltonian for the proton in the constituent valence quark Fock space ($|qqq\rangle$) and solve for its mass eigenstates and LFWFs. Parameters in this Hamiltonian have been fitted to generate the proton mass and the electromagnetic properties of the nucleon~\cite{Mondal:2019jdg, Xu:2021wwj}. The resulting LFWFs have successfully been used to investigate various proton properties, such as electromagnetic and axial form factors, radii, PDFs, GPDs, TMDs, and angular momentum distributions, etc.,~\cite{Mondal:2019jdg, Xu:2021wwj, Liu:2022fvl, Hu:2022ctr, Liu:2024umn, Kaur:2023lun}. Here, we extend those investigations to study both the unpolarized and longitudinally polarized quark DPDs of the proton.

	We structure the paper as follows. Section~\ref{sec:hami} introduces the BLFQ framework and outlines our Hamiltonian, which we diagonalize to derive the LFWFs. In section~\ref{DPDs_WFs}, we present the overlap representation for both unpolarized and polarized DPDs. Section~\ref{sec:results} presents our numerical results obtained using the resulting LFWFs. Finally, we present our summary and outlooks in section ~\ref{sec:conclusion}.
	\section{ BASIS LIGHT-FRONT QUANTIZATION}\label{sec:hami}
	In BLFQ, we aim to solve the light-front eigenvalue problem, expressed as:
	$H_{{\rm LC}} | \psi_{\rm h} \rangle = M^2_{\rm h} | \psi_{\rm h} \rangle,$
	where the light-front Hamiltonian, $H_{{\rm LC}} = (P^+ P^- - P^2_{\perp})$, incorporates the longitudinal momentum\footnote{We adopt a convention for the light-front variables~\cite{Harindranath:1996hq} such that for a given four-vector $x^\mu$, we have $x^{\pm} = x^0 \pm x^3$ and the remaining transverse degrees of freedom are denoted by $\vec{x}_{\perp}=(x_1,x_2)$.} $P^+$, the light-front energy $P^-$, and the transverse momentum $\vec{P}_\perp$. Solving this equation with a suitable matrix representation for $H_{{\rm LC}}$ yields the invariant-mass spectrum $M_{\rm h}$ and light-front state vectors $| \psi_{\rm h} \rangle$, which provide the LFWFs when expanded in momentum space.
	
	The expansion of the state $| \psi_{\rm h} \rangle$ within one Fock sector utilizes a complete set of orthonormal basis states. The BLFQ basis comprises transverse two-dimensional (2D) harmonic-oscillator (HO) modes, discretized longitudinal plane waves, and the light-front helicity eigenstates.    
	Thus, basis states are characterized by four quantum numbers $\alpha = (k,n,m,s)$, where $n$ and $m$ are the principal and orbital angular quantum numbers in the transverse plane, respectively, $s$ denotes the light-front helicity, and $k$ represents the longitudinal momentum.

	The orthonormalized 2D-HO basis functions in momentum space are expressed as ~\cite{Vary:2009gt,Zhao:2013cma}:
	\begin{align}
		\phi_{n,m}(\vec{k}_{\perp};b)
		=&\frac{\sqrt{2}}{b(2\pi)^{\frac{3}{2}}}\sqrt{\frac{n!}{(n+|m|)!}}e^{-\vec{k}_{\perp}^2/(2b^2)}\nonumber\\
		&\times\left(\frac{|\vec{k}_{\perp}|}{b}\right)^{|m|}L^{|m|}_{n}\left(\frac{\vec{k}_{\perp}^2}{b^2}\right)e^{im\theta},\label{ho}
	\end{align}
	with $b$ as the HO basis scale parameter, and $L^{m}_{n}$ the generalized Laguerre polynomials. To obtain a finite basis as a requirement from numerical calculations, we further implement the following truncation in the transverse plane $\sum_i (2n_i + |m_i|+1) \le N_{\mathrm{max}}$, where $N_{\mathrm{max}}$ is the total number of oscillator quanta above the minimum required by the Pauli principle.
	
	Longitudinal modes are discretized by imposing periodic or antiperiodic boundary conditions for bosons and fermions, respectively, within a box of length $2L$, leading to the discretized plane wave in coordinate space:
	$\psi_{k}(x^-) = \frac{1}{2L}~e^{i \frac{\pi}{L}k x^-}$, with quantum number $k$ as integer (bosons) or
	half-integer (fermions). Total longitudinal momentum $P^+$ is conserved for all basis states with $P^+ = \pi K/L$ and the longitudinal momentum fraction is $x_i \equiv p_i^+/P^+ = k_i/K$. 
	
	Due to the use of the single-particle basis, the solved eigenstates also contain the center-of-mass (CM) motion. To avoid the contamination of the CM motion, a Lagrange multiplier is added to adjust the Hamiltonian: $H' = H_{\rm LC} + \lambda_{\rm L} (H_{\rm CM} - 2b^2 I),$	where the CM motion is governed by a HO potential~\cite{Wiecki:2014ola}: $H_{\rm CM}=\left(\sum_i \vec{k}_{i\perp}\right)^2+b^4\left(\sum_i x_i  \vec{r}_{i\perp}\right)^2.$	The introduction of the Lagrange multiplier, together with the use of the HO basis and the $N_{\mathrm{max}}$ truncation allows for the factorization of the CM and intrinsic components of the LFWFs.

	This study focuses exclusively on the leading Fock sector to set an initial baseline for the valence quark contributions. With this Fock truncation, we introduce an effective light-front Hamiltonian given by
	\begin{align}
		H_{\mathrm{eff}} = \sum_i \frac{\vec{k}^2_{i\perp} + m^2_i}{x_i} + \frac{1}{2}\sum_{i,j} V_{ij}^{(\mathrm{CON})} + \frac{1}{2} \sum_{i,j} V_{ij}^{(\mathrm{OGE})} ,
		\label{hamieff}
	\end{align}
	as an initial model for  the first-principles $H_{\rm LC}$ appropriate to the Fock sector truncation. In the effective light-front Hamiltonian $H_{\rm eff}$, the conditions $\sum_i x_i =1$ and $\sum_i \vec{k}_{i\perp} =\vec{P}_\perp$ are enforced for momentum conservations, $m_i$ represents the constituent mass of the quark, and indices $i$ and $j$ identify different partons within the Fock sector. The first component of the Hamiltonian is the kinetic energy. The second and third terms are the confinement potential (CON) and the one-gluon exchange interaction (OGE) respectively. 
	
	The confinement potentials are designed to act in both transverse and longitudinal directions. Specifically, the transverse confinement potential draws inspiration from the soft-wall (SW) holographic model \cite{Brodsky:2014yha}, which is extended to a many-body framework as,
	\begin{align}
		V_{ij}^{\mathrm{SW}} = \kappa^4 x_i x_j (\vec{r}_{i\perp}-\vec{r}_{j\perp})^2.
	\end{align}
	For the longitudinal confinement potential \cite{Li:2015zda}, the many-body generalized form can be written as follows:
	\begin{align}
		V_{i,j}^{\mathrm{L}} = \frac{\kappa^4}{(m_i + m_j)^2} \partial_{x_i}(x_i x_j \partial_{x_j})	.
	\end{align}
	The strength of the confining potential is represented by the parameter $\kappa$. Consequently, the complete confinement potential is expressed as $V_{ij}^{(\mathrm{CON})} = V_{ij}^{\mathrm{SW}} + V_{ij}^{\mathrm{L}}$.
	
	The OGE interaction is expressed as,
	\begin{align}
		V_{ij}^{\mathrm{OGE}} = \frac{4\pi \alpha_s C_F}{Q^2_{ij}} \bar{u}_{s'_i}(k'_i)\gamma^{\mu}u_{s_i}(k_i)\bar{u}_{s'_j}(k'_j)\gamma_{\mu}u_{sj}(k_j).
	\end{align}
	Here $Q^2_{ij} = -(1/2)(k'_i - k_i)^2 - (1/2)(k'_j - k_j)^2$ represents the squared average momentum transfer, where $k_i(k'_i)$ and $k_j(k'_j)$ are the initial (final) momenta of particles $i$ and $j$, respectively. $u_{s_i}(k_i)$ denotes the Dirac spinor for a particle with momentum $k_i$ and spin $s_i$. The $\alpha_s$ indicates the strength of the OGE interaction. The negative value of color factor $ C_F = -2/3 $ signifies that the OGE potential is attractive.
	
	Diagonalizing the above light-front Hamiltonian using the BLFQ basis yields eigenvalues that define the mass spectrum, together with the associated eigenvectors that describe the inner structures of the systems. The lowest mass eigenstate is identified as the proton state. The proton state with momentum $P$ and helicity $\Lambda$  within the valence Fock sector can be expressed using three-particle LFWFs as,
	\begin{align}
		\label{protonstate}
		\mid P,\Lambda \rangle =& \int 	\left[{\rm d}\mathcal{X} \,{\rm d}\mathcal{P}_\perp\right] \nonumber\\
		&\times\Psi^{\Lambda}_{\{x_i,\vec{k}_{i\perp},s_i\}} \Bigl\vert \left\{ x_i P^+, \vec{k}_{i\perp} + x_i\vec{P}_{\perp}, s_i \right\} \Bigl\rangle,
	\end{align}
	where the shorthand notation used for the integration measure is written as,
	\begin{align}\label{notation}
		\left[{\rm d}\mathcal{X} \,{\rm d}\mathcal{P}_\perp\right]=&\prod_{i=1}^3 \left[\frac{{\rm d}x_i{\rm d}^2 \vec{k}_{i\perp}}{16\pi^3}\right]\nonumber\\&\times 16 \pi^3 \delta \left(1-\sum_{i=1}^{3} x_i\right) \delta^2 \left(\sum_{i=1}^{3}\vec{P}_\perp-\vec{k}_{i\perp}\right) .  
	\end{align}
	
	The LFWFs are expressed in terms of the basis functions as:
	\begin{align}\label{wavefunctions}
		\Psi^{\Lambda}_{\{x_i,\vec{k}_{i\perp},s_i\}}=&\sum_{\{k_i,n_i,m_i\}} \psi^{\Lambda}_{\{k_{i},n_{i},m_{i},s_i\}}\\\nonumber 
		&(\prod_i \delta_{x_i}^{\frac{k_i}{K}})(\prod_i \phi_{n_i,m_i}(\vec{k}_{i\perp};b)),
	\end{align}
	where $\psi^{\Lambda}_{\{k_{i},n_{i},m_{i},s_i\}}= \langle P,\Lambda | \{k_i,n_i,m_i,s_i\} \rangle$ represents the proton's LFWF in the BLFQ basis, obtained from diagonalizing the light-front Hamiltonian. 
	Here, $\phi_{n_i,m_i}(\vec{k}_{i\perp};b)$ is the orthonormalized 2D harmonic oscillator defined in Eq.~\ref{ho}, and $\delta_{x_i}^{k_i/K}$ stands for the Kronecker delta.

	The parameters within the effective Hamiltonian are tuned to accurately reflect the nucleon mass and its electromagnetic characteristics~\cite{Xu:2021wwj}. The model LFWFs have been utilized to explore a diverse range of nucleon properties, such as electromagnetic and axial form factors, radii, PDFs, quark helicity asymmetries, GPDs, TMDs and angular momentum distributions, etc.~\cite{Mondal:2019jdg, Xu:2021wwj, Liu:2022fvl, Hu:2022ctr, Liu:2024umn, Kaur:2023lun}. These studies have achieved notable success, demonstrating the comprehensive applicability of the current model in calculating hadron observables by solving bound state eigenvalue problems.
	\section{Double parton distributions}\label{DPDs_WFs}
	The DPDs of a hadron provide a probabilistic framework that describes the likelihood of encountering two partons, each with specified polarization and momentum fractions $x_i$, situated at a specific relative transverse distance $\vec{y}_\perp$. The DPDs of quarks in an unpolarized proton are defined by a matrix element involving two operators \cite{Diehl:2011tt, Diehl:2011yj},
	\begin{align}\label{dpd_eq}	
		F_{a_1a_2}(x_1,x_2,y_\perp )&=2p^+ \int dy^- \int \frac{dz_1^-}{2\pi}\frac{dz_2^-}{2\pi}e^{i(x_1z_1^-+x_2z_2^-)p^+}\nonumber\\
		&\times\left\langle P,\Lambda\right |\mathcal{O}_{a_1}(y,z_{1})\mathcal{O}_{a_2} (0,z_{2})\left |P,\Lambda\right \rangle.
	\end{align}
	The operators appearing in Eq.~\eqref{dpd_eq} are defined as,	
	
	\begin{align}\label{quark_op}		
		\mathcal{O}_{a}(y,z)=\left. \bar{\psi}\left(y-\frac{z}{2}\right)\Gamma_{a}\psi\left(y+\frac{z}{2}\right)\right\rvert_{z^+=y^+=0,\vec{z}_{\perp}=0},
	\end{align}	
	where $a$ specifies the polarization of the quark, which is determined by the spin projections through
	\begin{align}\label{quark_op2}		
		\Gamma_q=\frac{1}{2}\gamma^+, \hspace{0.4cm}	\Gamma_{\Delta_q}=\frac{1}{2}\gamma^+\gamma_5,	 \hspace{0.4cm}	\Gamma_{\delta_q}^j=\frac{1}{2}i\sigma^{j+}\gamma_5.
	\end{align}	
	Here $j=(1,2)$ and the notations appearing in the subscript $q$, $\Delta_q$ and $\delta_q$ refer to unpolarized, longitudinally polarized, and transversely polarized quark, respectively. We choose the light-cone gauge, and thus the gauge link in this bilocal operator is reduced to unity.

	Substituting the proton state Eq.~\eqref{protonstate} into Eq.~\eqref{dpd_eq}, from the  unpolarized operator corresponding to $\Gamma_q=\frac{1}{2}\gamma^+$, we obtain the following expression for the unpolarized quark DPDs in terms of the overlap of LFWFs,
	\begin{align}\label{unpolarized_quarks_DPDs}	
		F_{q_1q_2}&(x_1,x_2,\vec{y}_\perp)=\sum_{s_i} \int {\rm d}^2\vec{k}_{1 \perp}{\rm d}^2\vec{k}_{2 \perp}{\rm d}^2\vec{k}_{1 \perp}^{\prime}\nonumber\\
		&\times\Psi^{\uparrow *}_{\{x_i,\vec{k}{_{i \perp}^{\prime}},s_i\}}\Psi^{\uparrow}_{\{x_i,\vec{k}{_{i \perp}},s_i\}} e^{i(\vec{k}_{1 \perp}^{\prime}-\vec{k}_{1 \perp}).\vec{y}_\perp},
	\end{align}
	where for struck quarks we have $\vec{k}{_{2 \perp}^{\prime}}=\vec{k}{_{1 \perp}}+\vec{k}{_{2 \perp}}-\vec{k}{_{1 \perp}^{\prime}}$ and $\vec{k}_{3\perp}^\prime=\vec{k}_{3\perp}=-(\vec{k}_{1\perp}+\vec{k}_{2\perp})$ for the spectator.  $x_1$ and $x_2$ are the longitudinal moment fractions of the struck quarks, and $\vec{y}_\perp$ is the relative distance in the transverse plane between the struck quarks. 
	
	Integrating Eq. (12) over the transverse relative distance $\vec{y}_{\perp}$ produces a delta function $\delta^2(\vec{k}_{1\perp}-\vec{k}'_{1\perp})$, which removes the integration over $\vec{k}'_{1\perp}$. Subsequently, by integrating over the momentum fraction $x_1 (x_2)$, one explicitly recovers the standard overlap representation of the unpolarized PDF for the parton $q_2 (q_1)$, within the same Fock-sector truncation scheme (see, e.g., Eq. (6) in Ref~\cite{Mondal:2019jdg}). We have numerically verified this integral relationship, confirming excellent agreement. This result provides strong evidence for the internal consistency and probabilistic interpretation of the DPDs computed within our BLFQ-based approach.
	
	Similarly, the overlap representation for the DPDs of two longitudinally polarized quarks corresponding to the operator with $\Gamma_{\Delta_q}=\frac{1}{2}\gamma^+\gamma_5$ can be written as:
	\begin{align}\label{polarized_quarks_DPDs}	
		F_{\Delta q_1\Delta q_2}&(x_1,x_2,\vec{y}_\perp)=\sum_{s_i} (4 s_1 \cdot s_2) \int {\rm d}^2\vec{k}_{1 \perp}{\rm d}^2\vec{k}_{2 \perp}{\rm d}^2\vec{k}_{1 \perp}^{\prime}\nonumber\\
		&\times\Psi^{\uparrow *}_{\{x_i,\vec{k}{_{i \perp}^{\prime}},s_i\}}\Psi^{\uparrow}_{\{x_i,\vec{k}{_{i \perp}},s_i\}} e^{i(\vec{k}_{1 \perp}^{\prime}-\vec{k}_{1 \perp}). \vec{y}_\perp},
	\end{align}
	where $s_1$ and $s_2$ are the light-front helicities of the two struck quarks.
	A similar procedure applied directly to Eq. \ref{polarized_quarks_DPDs}, which describes the distribution of two longitudinally polarized partons, does not yield the overlap representation of the standard longitudinally polarized PDF $g_1$. The correct sum rule relevant in this context relates instead the polarized PDF of one parton to the polarized-unpolarized DPD, that is, the distribution of a longitudinally polarized parton combined with an unpolarized parton. Therefore, a direct integration check as described after Eq.  \ref{unpolarized_quarks_DPDs} does not apply here.

	\section{Numerical Results}
	\label{sec:results}
	The LFWFs of the proton have been obtained in the BLFQ framework with the basis truncation $N_{\rm max}=10$ and $K=16.5$, along with model parameters $\{m_{\rm u/KE},~m_{\rm d/KE},~m_{\rm q/OGE},~\kappa,~\alpha_s,~b\}=\{0.300~{\rm GeV},~0.301~{\rm GeV},~0.200~{\rm GeV},~0.34~{\rm GeV},~1.1\pm 0.1,~~0.6~{\rm GeV}\}$, where $m_{q/{\rm K.E.}}$ and $m_{q/{\rm OGE}}$ correspond to the quark masses in kinetic energy and OGE interaction terms, and $b$ is the HO scale parameter.
	
	The parameters in our model are determined to reproduce the known nucleon mass and electromagnetic form factors~\cite{Mondal:2019jdg, Xu:2021wwj}. 
	The LFWFs of the valence Fock sector with these parameters, with an implicit model scale of $\mu^2_0=0.195 \pm 0.020$ GeV$^2$~\cite{Mondal:2019jdg, Xu:2021wwj}, are utilized to derive physical observables and distribution functions of the valence quarks within the proton. Our focus in this work is on investigating the DPDs of the proton using these LFWFs. We insert the valence LFWFs given by Eq.~(\ref{wavefunctions}) into Eqs.~(\ref{unpolarized_quarks_DPDs}) and (\ref{polarized_quarks_DPDs}) to compute the valence quark DPDs inside the proton.

	\subsection{Unpolarized quark DPDs $F_{q_1q_2}(x_1,x_2,\vec{y}_\perp)$}
	The unpolarized quark DPDs, denoted as $F_{q_1q_2}(x_1,x_2,\vec{y}_\perp)$, quantify the likelihood of identifying two quarks within a hadron, carrying momentum fractions $x_1$ and $x_2$, and separated by a transverse distance $\vec{y}_\perp$. This probabilistic description is restricted to the physically meaningful kinematic region, more specifically where the sum of the momentum fractions does not exceed unity, $x_1 + x_2 < 1$. This constraint ensures that the calculations obtain support from the chosen basis configurations of BLFQ.
	
	\begin{figure}[tph]
		\centering
		\includegraphics[width=0.5\textwidth]{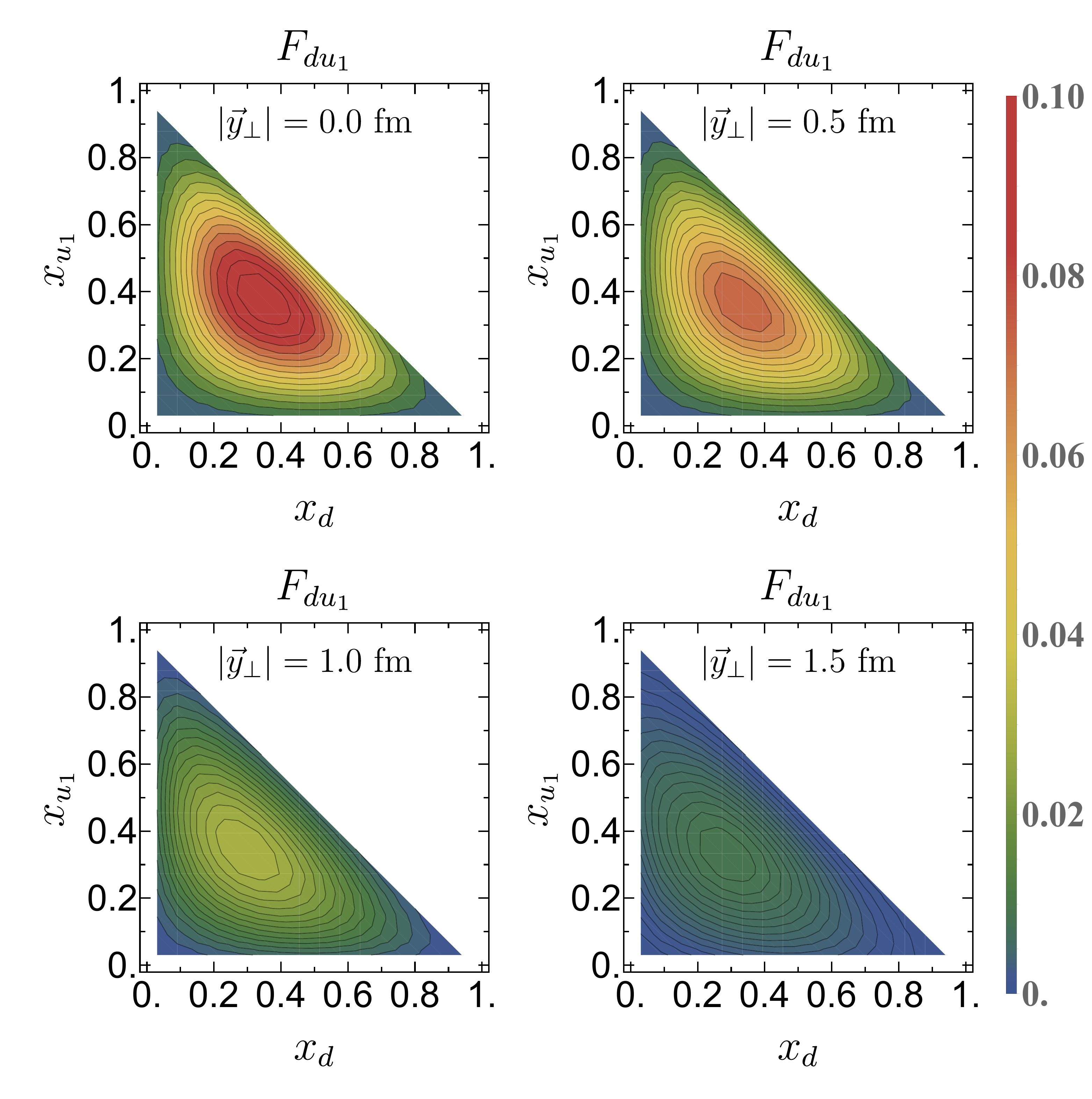}
		\caption{The unpolarized DPDs $F_{du_1}(x_d,x_{u_1},\vec{y}_\perp)$ as functions of $x_{d}$ and $x_{u_1}$ for different values of $|\vec{y}_\perp|=\{0,\,0.5,\,1.0,\,1.5\}$ fm.} 
	\label{fig1}
\end{figure}

\begin{figure}[tph]
	\centering
	\includegraphics[width=0.5\textwidth]{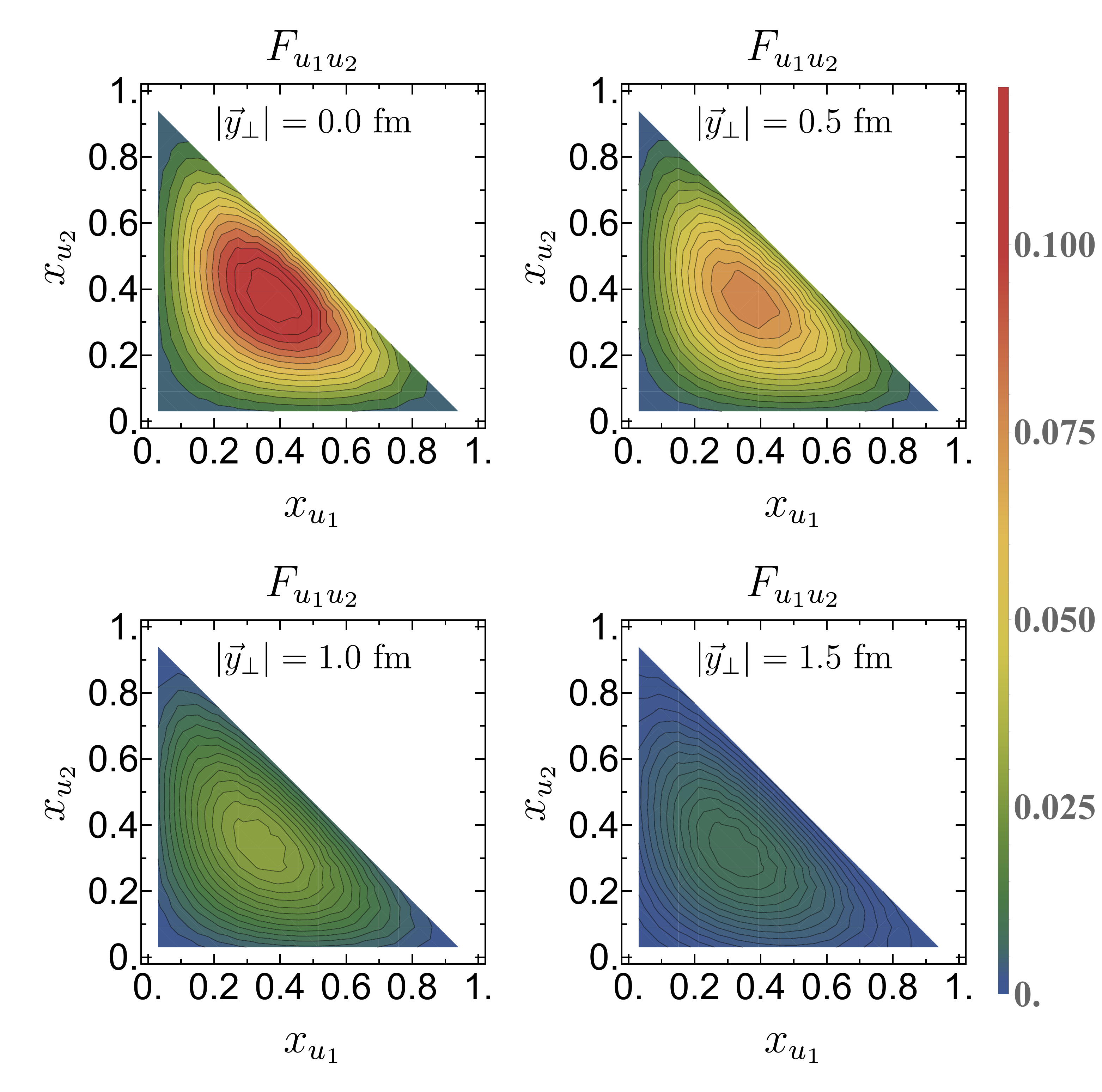}
	\caption{The unpolarized DPDs $F_{u_1 u_2}(x_{u_1},x_{u_2},\vec{y}_\perp)$ as functions of $x_{u_1}$ and $x_{u_2}$ for different values of $|\vec{y}_\perp|=\{0,\,0.5,\,1.0,\,1.5\}$ fm.}
\label{fig2}
\end{figure}

\begin{table}[ht]
\caption{The peak positions and values of unpolarized DPDs $F_{d u_1}$ and $F_{u_1 u_2}$ in Fig.\ref{fig1} and \ref{fig2}.}
\centering
\begin{tabular}[t]{c|c|c|c|c}
	\toprule
	$|\vec{y}_\perp|/\mathrm{fm}$ ~~ & ~~$(x_d,x_{u_1})$~~  &~values ~~ & ~~$(x_{u_1},x_{u_2})$~~ &~values\\
	\hline
	$0 $   &~~(0.347,0.376) &~0.102 &~~(0.377,0.377) &~0.122 \\
	$0.5 $ &~~(0.333,0.365) &~0.076 &~~(0.366,0.366) &~0.081 \\
	$1.0 $ &~~(0.296,0.343) &~0.030 &~~(0.341,0.341) &~0.029 \\
	$1.5 $ &~~(0.287,0.327) &~0.009 &~~(0.321,0.321) &~0.009 \\
	\hline\hline
\end{tabular}
\label{table_peak1}
\end{table}

Figure~\ref{fig1} shows the contour plots depicting the unpolarized DPDs for a down quark ($d$) and an up quark ($u_1$)~\footnote{Here, the subscript 1 and 2 for $u$ are just labels since in our calculation those two up quarks are indistinguishable. The only reason to include those subscripts is to stress that here we show only half of the DPDs between $d$ and $u$ quark. Similarly for $F_{u_1 u_2}$ later.} within the proton, plotted against their longitudinal momentum fractions $x_{d}$ and $x_{u_1}$. We show the distributions for various transverse distances $|\vec{y}_\perp|$, ranging from $|\vec{y}_\perp| = 0~\mathrm{fm}$ to $|\vec{y}_\perp| = 1.5~\mathrm{fm}$.
For ease of visualization, the values corresponding to different colors change from one plot to the other as indicated by the color bars to the right of each plot.
Fig.~\ref{fig1} shows that the unpolarized DPDs first increase and then decrease when $x_d +x_{u_1}$ increases, with the peak around $x_{d} \approx x_{u_1} \approx 0.3$. 
Table. \ref{table_peak1} explicitly shows the numerical locations of peaks in the unpolarized DPDs. As evident, the peak positions systematically shift toward lower values of $x$ as the transverse distance $|\vec{y}_{\perp}|$ between the quarks increases. 
This behavior arises from the weakening of partonic correlations at large transverse separation, where the system approaches a decorrelated regime and low-$x$ configurations become more prominent. Additionally, the table reveals a structural difference between the two flavor combinations. In the $F_{u_1 u_2}$ case, the peak consistently lies along the $x_{u_1} = x_{u_2}$ diagonal, reflecting the symmetry between the two identical $u$ quarks. In contrast, for the $F_{d u_1}$ case, the peak position is systematically off-diagonal, with $x_d < x_{u_1}$ at all $|\vec{y}_\perp|$. This asymmetry originates from the small but nonzero mass difference between the $u$ and $d$ quarks in our model, which breaks exact flavor symmetry. As a result, the $d$ quark consistently carries slightly less longitudinal momentum than the $u$ quark, leading to a peak location that lies below the $x_d = x_{u_1}$ diagonal. This asymmetry persists across all transverse separations shown in Fig.~\ref{fig1} and is quantitatively supported by the values in Table~\ref{table_peak1}.


Figure~\ref{fig2}, which focuses on the interaction between two up quarks $u_1$ and $u_2$, exhibits trends similar to that observed in Fig.~\ref{fig1}. However, a subtle shift in the peak positions between the two figures is observed which is more explicit in the large $|\vec{y}_\perp|$ case. 
The slightly asymmetric patterns for distinguishable ($u, d$) quarks seen in Fig.~\ref{fig1} that result from our small (u,d) quark mass splitting are replaced by the symmetric patterns seen in Fig.~\ref{fig2} (and corresponding results in Table. \ref{table_peak1}) that reflect the inherent indistinguishability of the ($u_1, u_2$) quarks.
This variation is attributed to the slight numerical differences in the masses (1 MeV) between the up and down quarks in our calculations, which breaks the isospin symmetry.
We also observe that for a fixed $|\vec{y}_\perp|$, the distribution $F_{u_1u_2}$ has a larger magnitude than that of the $F_{du_1}$, which signifies a stronger correlation between the two $u$ quarks than that between the $d$ and $u$ quarks in the proton.


\begin{figure}[tph]
\centering
\includegraphics[width=0.5\textwidth]{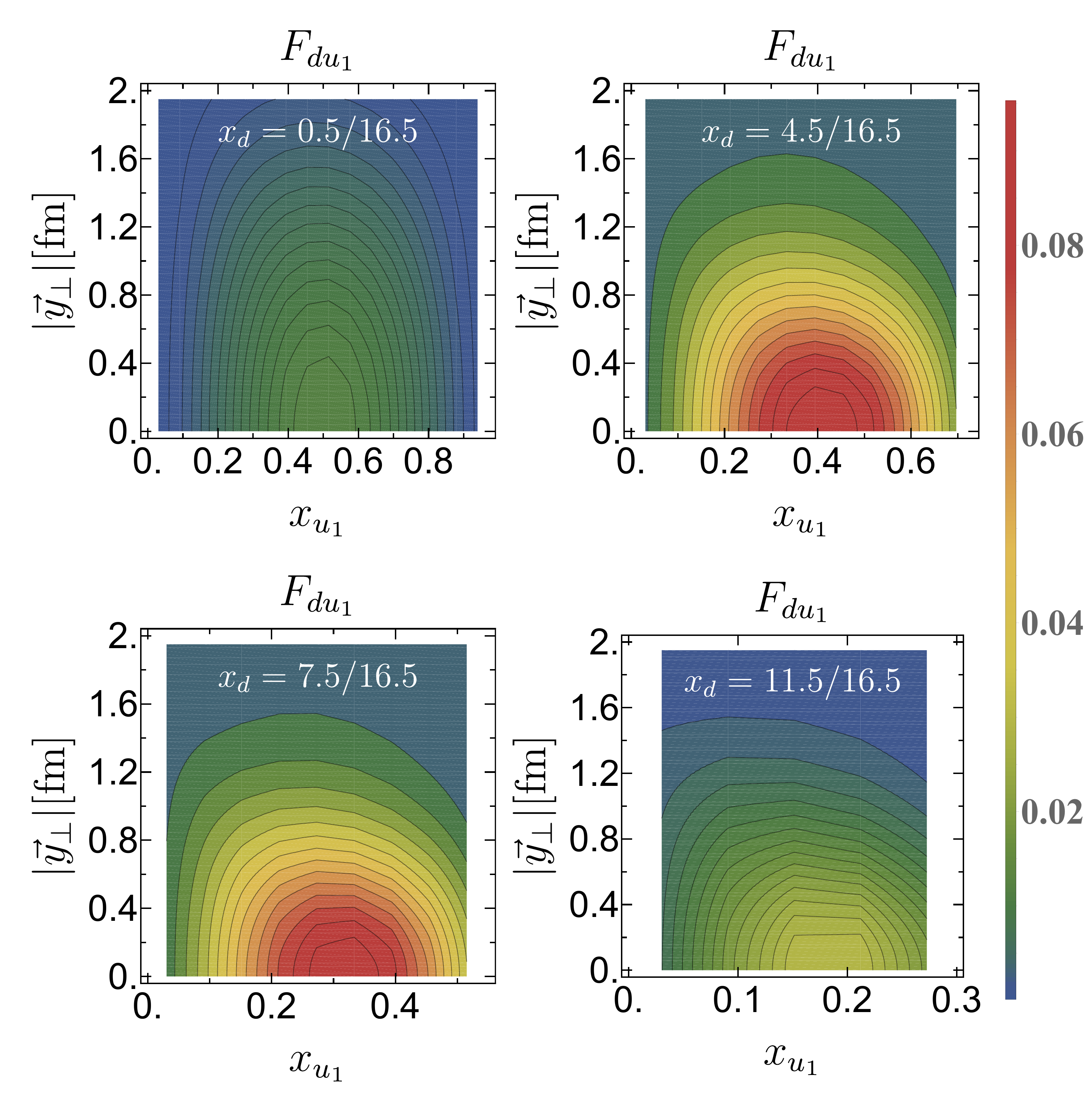}
\caption{The unpolarized DPDs $F_{du_1}(x_d,x_{u_1},\vec{y}_\perp)$ as functions of $x_{u_1}$ and $
	|\vec{y}_\perp|$ for different values of $x_d=\{0.5,\,4.5,\,7.5,\,11.5\}/16.5$.}
\label{fig3}
\end{figure}

\begin{figure}[tph]
\centering
\includegraphics[width=0.5\textwidth]{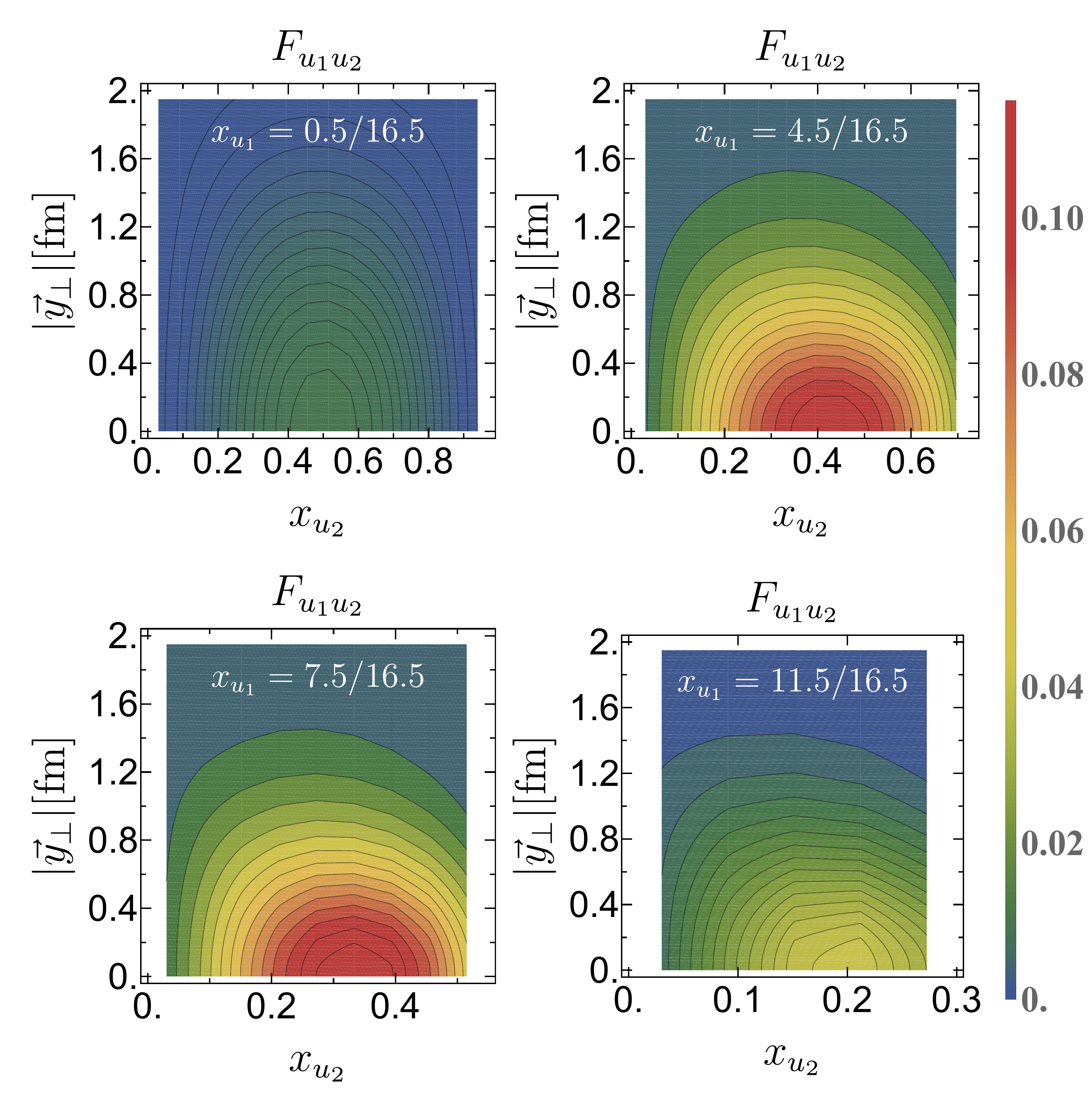}
\caption{The unpolarized DPDs $F_{u_1u_2}(x_{u_1}x_{u_2},\vec{y}_\perp)$ as functions of $x_{u_2}$ and $
	|\vec{y}_\perp|$ for different values of $x_{u_1}=\{0.5,\,4.5,\,7.5,\,11.5\}/16.5$.}
\label{fig4}
\end{figure}

In  Fig.~\ref{fig3}, we show the contour plot of the unpolarized DPDs $F_{du_1}$ as functions of $x_{u_1}$ and $\vec{y}_\perp$ with fixed values of $x_d=\{0.5,\,4.5,\,7.5,\,11.5\}/16.5$.
The DPDs achieve their maximum values when the relative distance between the two quarks is zero. As $x_{d}$ increases the peak value of the distribution first rises and then falls off. We observe that the distribution exhibits an approximate symmetry over $x_{u_1}$ when  $x_d$ is at its smallest value. However, this symmetry is broken as $x_d$ increases. Similar observations can also be noticed in Fig.~\ref{fig4}, where the unpolarized DPDs $F_{u_1u_2}$ are shown as functions of $x_{u_2}$ and $\vec{y}_\perp$ with fixed values of $x_{u_1}$.
\begin{figure}[tph]
\centering
\includegraphics[width=0.45\textwidth]{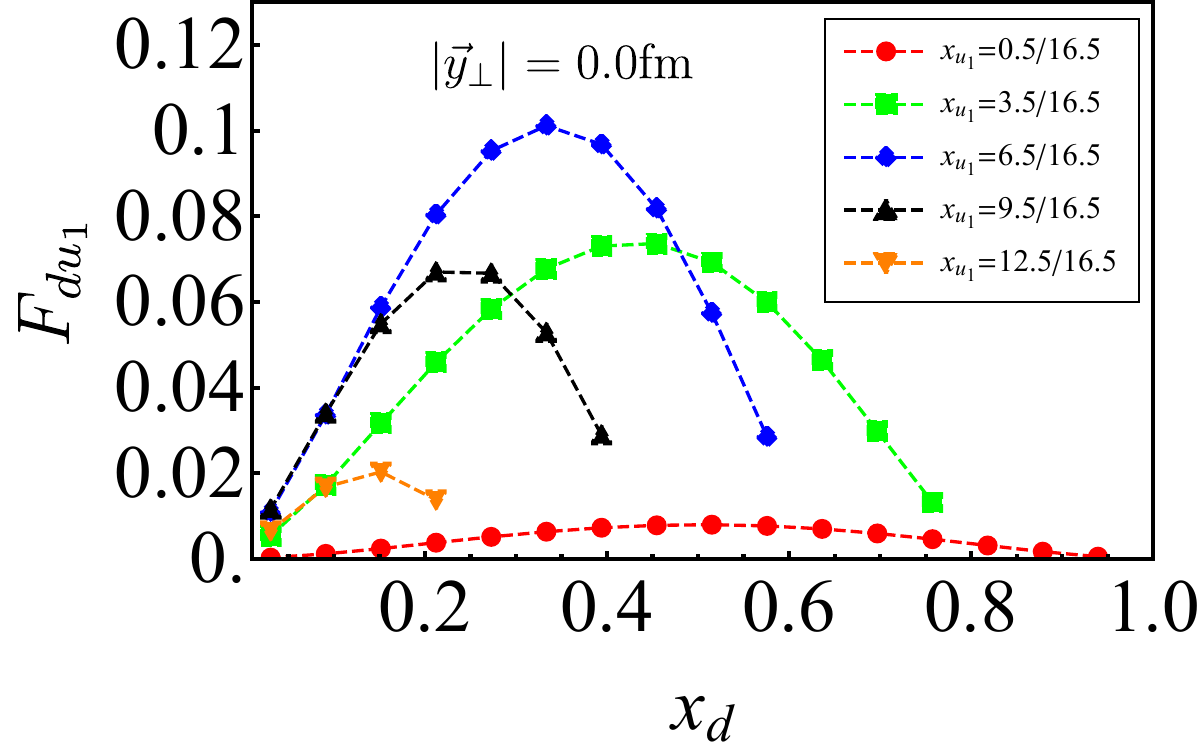}
\includegraphics[width=0.45\textwidth]{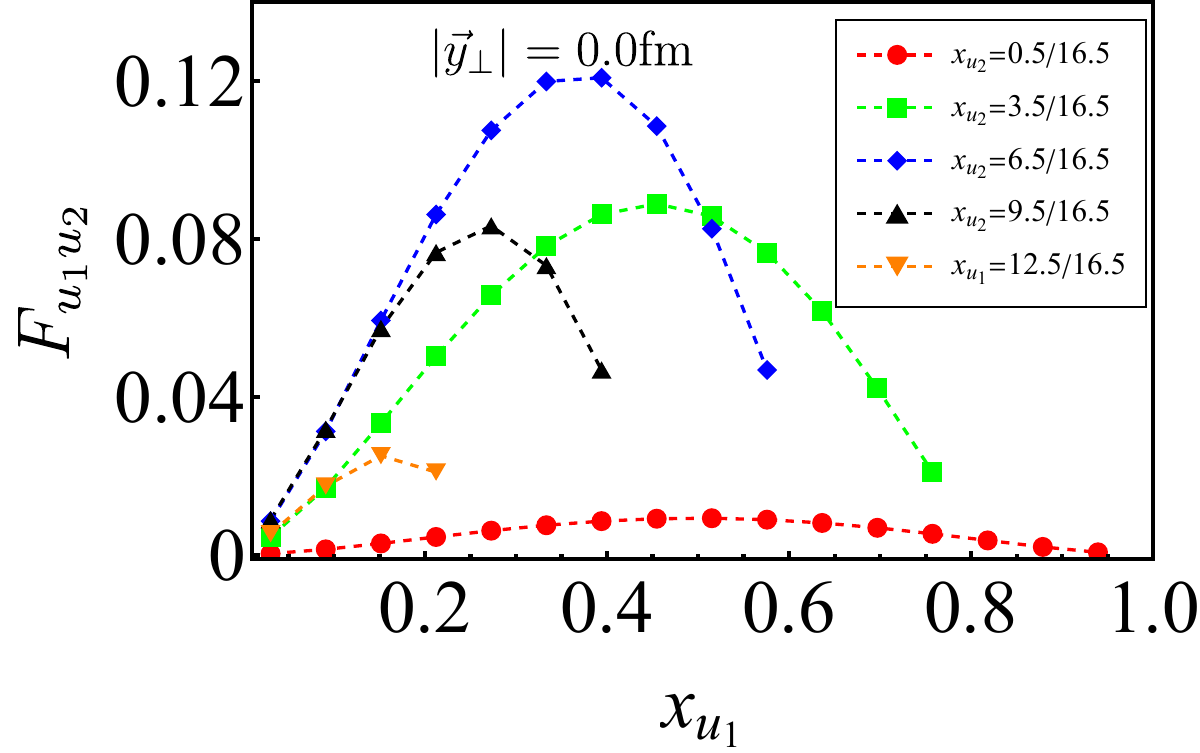}
\caption{The unpolarized DPDs,  $F_{du_1}(x_d,x_{u_1},\vec{y}_\perp)$ (upper panel)  and  $F_{u_1u_2}(x_{u_1},x_{u_2},\vec{y}_\perp)$ (lower panel), as functions of $x_d (x_{u_1})$ for different values of $x_{u_1} (x_{u_2})=\{0.5,\,3.5,\,6.5,\,9.5,\,12.5\}/16.5$ and fixed $|\vec{y}_\perp|=0$ fm.}
\label{fig5}
\end{figure}

In Fig.~\ref{fig5}, we show the 2D plots of the unpolarized DPDs $F_{d u_1} (F_{u_1 u_2})$ as functions of $ x_{d}~(x_{u_1})$ for fixed values of $x_{u_1}(x_{u_2}) $ at the transverse distance of $|\vec{y}_\perp|= 0 ~\mathrm{fm}$. We observe that as the value of $x_{u_1}(x_{u_2})$ increases the peak of the curve shifts towards lower values of $ x_{d}~(x_{u_1})$ and simultaneously the magnitude of the peak increases first and then decreases. 

\begin{figure}[tph]
\centering
\includegraphics[width=0.45\textwidth]{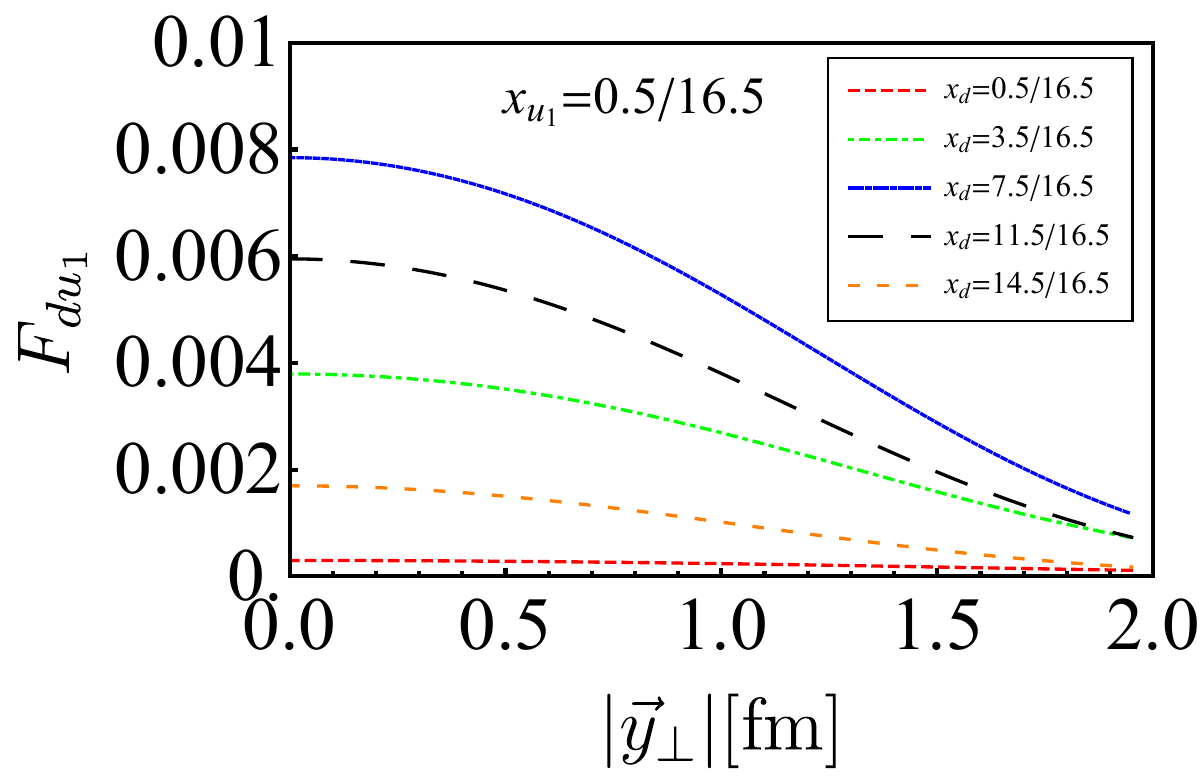}
\includegraphics[width=0.45\textwidth]{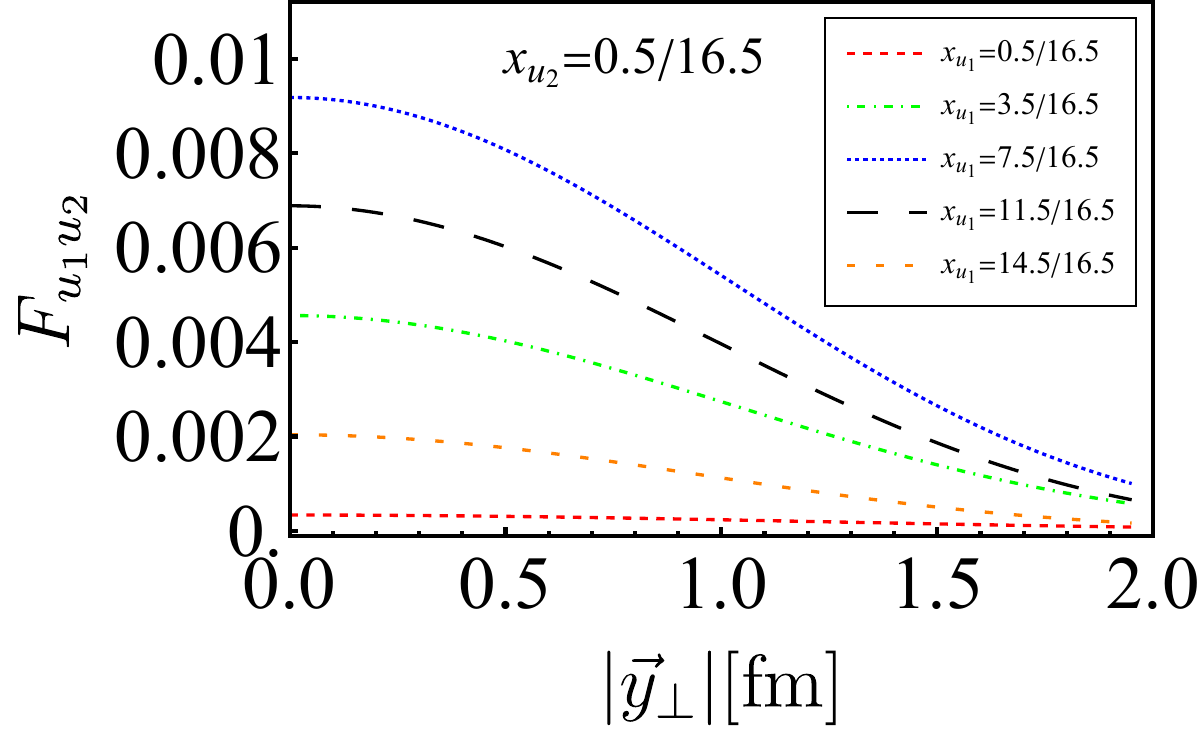}
\caption{The  unpolarized DPDs,  $F_{du_1}(x_d,x_{u_1},\vec{y}_\perp)$ (upper panel)  and  $F_{u_1u_2}(x_{u_1},x_{u_2},\vec{y}_\perp)$ (lower panel), as functions of $|\vec{y}_\perp|$ for different values of $x_d(x_{u_1})=\{0.5\,,3.5\,,7.5\,,11.5\,,14.5\}/16.5$ and fixed value of $x_{u_1}(x_{u_2})=0.5/16.5$.}
\label{fig6}
\end{figure}
In Fig.~\ref{fig6}, we plot the unpolarized DPDs $F_{d u_1} (F_{u_1 u_2})$ against the transverse distance $|\vec{y}_\perp|$ with fixed  $x_{u_1}(x_{u_2}) $ and different  $x_{d}~(x_{u_1})$, and we observe that as the distance between the two struck quarks increases, the probability of finding them decreases. With $x_{u_1}(x_{u_2})=0.5/16.5$, the magnitude of the distribution reaches its maximum around $x_{d}~(x_{u_1})  \approx 0.5$. Note that if we further increase the fixed value of $x_{u_1}(x_{u_2})$, the maximum of the distribution appears at $x_{d}~(x_{u_1})  < 0.5 $,  as can be seen in Figs.~\ref{fig3} and \ref{fig4}.

\subsection{Polarized quark DPDs $F_{\Delta q_1 \Delta q_2}(x_1,x_2,\vec{y}_\perp)$}

The polarized DPDs $F_{\Delta q_1 \Delta q_2}(x_1,x_2,\vec{y}_\perp)$ describe the difference between the probabilities of finding the two quarks with helicities both aligned and both anti-aligned to the proton helicity. Thus it gives a measure for the longitudinal quark polarization.
\begin{figure}[tph]
\centering
\includegraphics[width=0.5\textwidth]{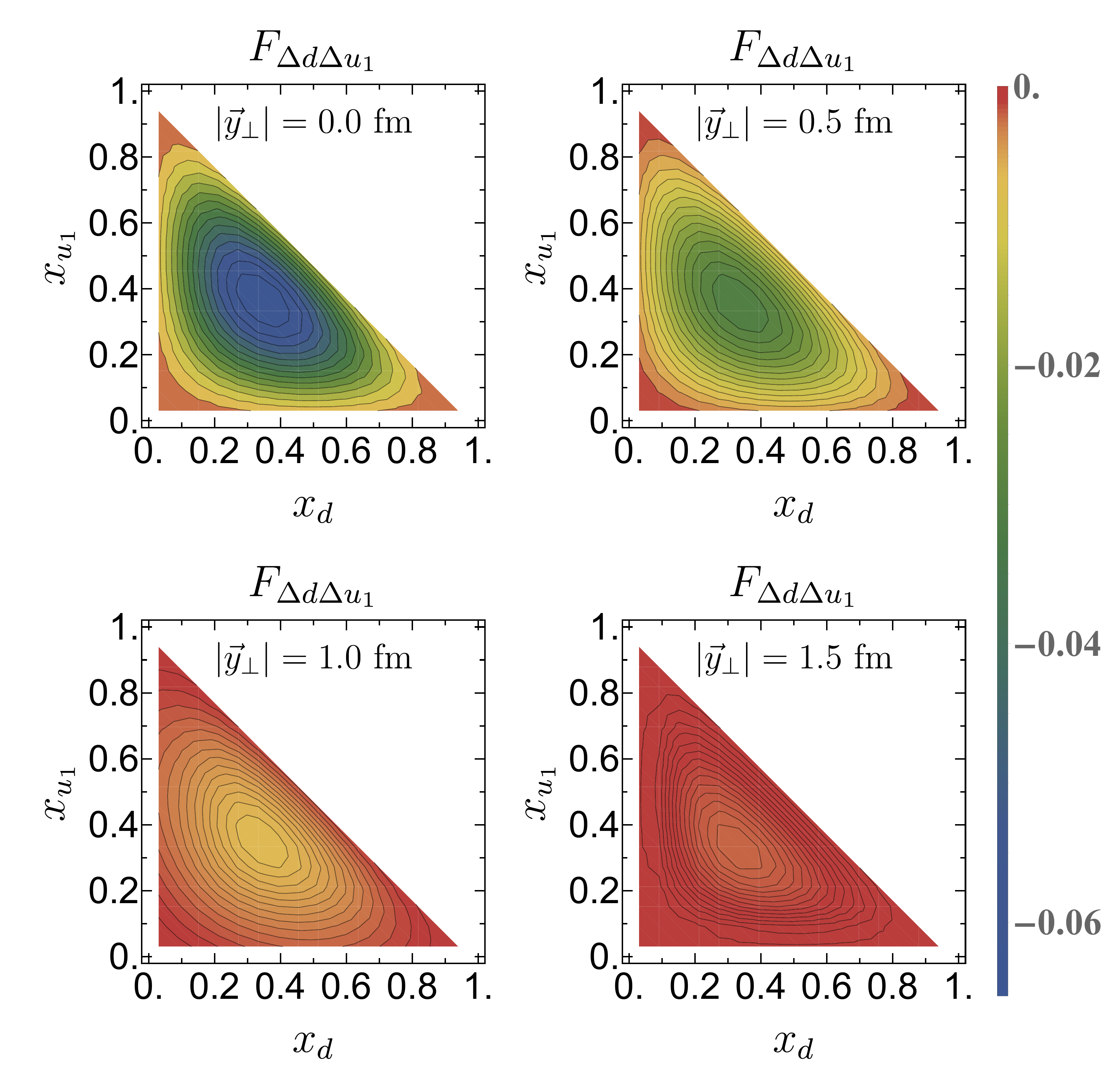}
\caption{The  polarized DPDs $F_{\Delta d \Delta u_1}(x_d,x_{u_1},\vec{y}_\perp)$ as functions of $x_{d}$ and $x_{u_1}$ for different values of $|\vec{y}_\perp|=\{0,\,0.5,\,1.0,\,1.5\}$ fm.}
\label{fig7}
\end{figure}
\begin{figure}[tph]
\centering
\includegraphics[width=0.5\textwidth]{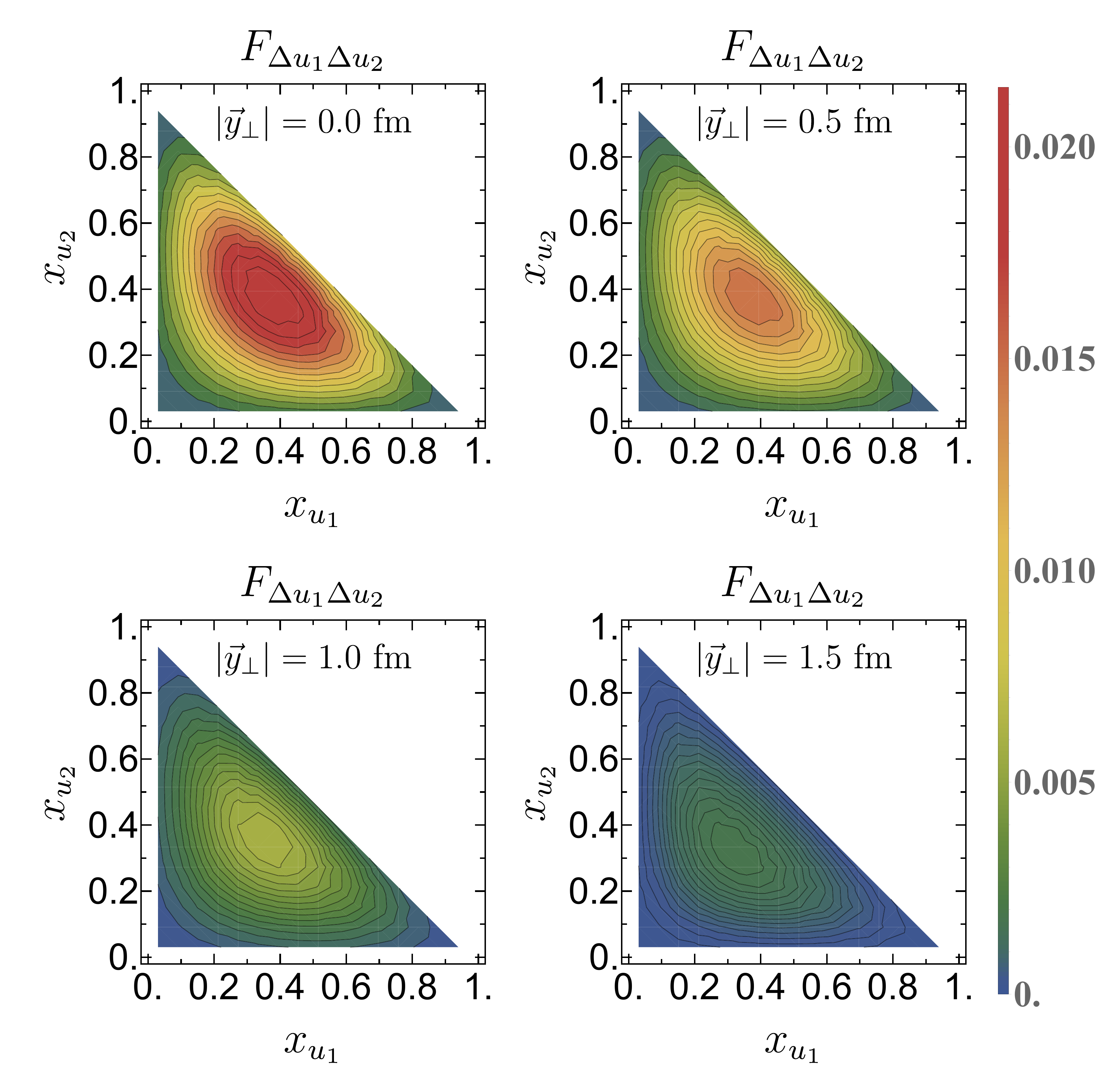}
\caption{The polarized DPDs $F_{\Delta u_1 \Delta u_2}(x_{u_1},x_{u_2},\vec{y}_\perp)$ as functions of $x_{u_1}$ and $x_{u_2}$ for different values of $|\vec{y}_\perp|=\{0,\,0.5,\,1.0,\,1.5\}$ fm.}
\label{fig8}
\end{figure}

\begin{table}[ht]
\caption{The peak positions and values of polarized DPDs $F_{\Delta d \Delta u_1}$ and $F_{\Delta u_1 \Delta u_2}$ in Fig.\ref{fig7} and \ref{fig8}.}
\centering
\begin{tabular}[t]{c|c|c|c|c}
	\toprule
	$|\vec{y}_\perp|/\mathrm{fm}$ ~~ & ~~$(x_d,x_{u_1})$~~  &~values ~~ & ~~$(x_{u_1},x_{u_2})$~~ &~values\\
	\hline
	$0 $   &~~(0.346,0.356) &~-0.066 &~~(0.377,0.377) &~0.022 \\
	$0.5 $ &~~(0.335,0.354) &~-0.032 &~~(0.374,0.374) &~0.015 \\
	$1.0 $ &~~(0.344,0.349) &~-0.007 &~~(0.356,0.356) &~0.006 \\
	$1.5 $ &~~(0.330,0.328) &~-0.002 &~~(0.322,0.322) &~0.002 \\
	\hline\hline
\end{tabular}
\label{table_peak2}
\end{table}

Figure~\ref{fig7} illustrates the polarized DPDs of $d$ and $u_1$ inside the proton as functions of their respective longitudinal momentum fractions $x_{d}$ and $x_{u_1}$ at transverse distance ranging from $|\vec{y}_\perp| = 0~\mathrm{fm}$ to $|\vec{y}_\perp| = 1.5~\mathrm{fm}$. Unlike the unpolarized distribution, $F_{d u_1}$, we find that when the struck quarks are longitudinally polarized, the double parton distributions relating up and down quarks are negative. 
We again find that the magnitudes of the polarized DPDs first increase and then decrease when $x_d +x_{u_1}$ increases, with the peak around $x_{d} \approx x_{u_1} \approx 0.3.$
A similar observation is also noticed for the polarized DPD of $u_1$ and $u_2$ quarks in Fig.~\ref{fig8}. However, in contrast with $F_{\Delta d \Delta u_1}$, $F_{\Delta {u_1} \Delta {u_2}}$ are positive, similar to the unpolarized distributions shown in Fig.~\ref{fig2}. Here we observe a clear effect of the longitudinal polarization from the different signs of $F_{\Delta d \Delta u_1}$ and $F_{\Delta u_1 \Delta u_2}$ compared with the same sign for the unpolarized case. 
Similar to the unpolarized DPDs, Table. \ref{table_peak2} presents numerical peak positions for the polarized DPDs, again indicating a clear trend of peak locations moving toward smaller $x$-values with increasing $|\vec{y}_{\perp}|$.

\begin{figure}[tph]
\centering
\includegraphics[width=0.5\textwidth]{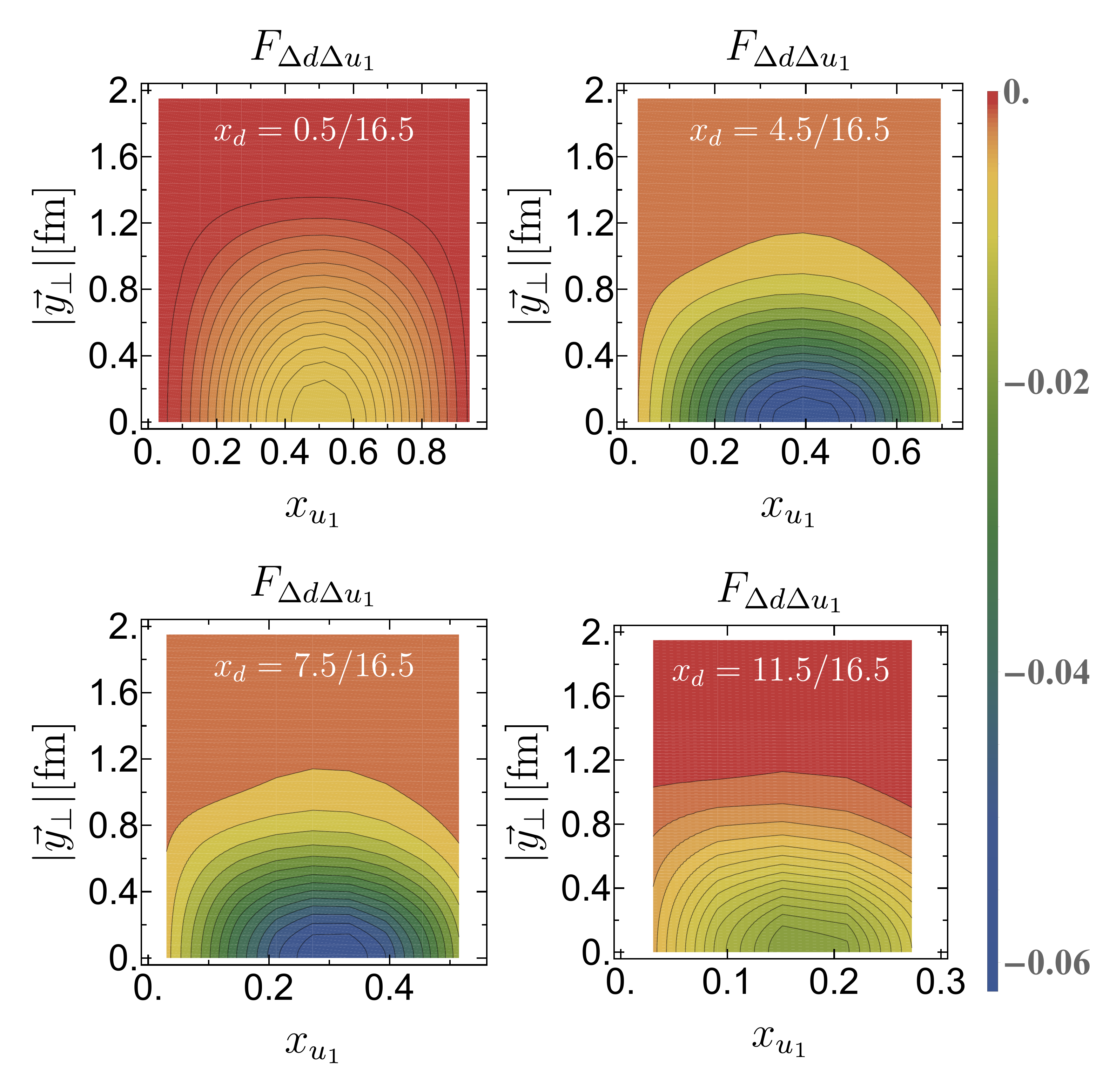}
\caption{The polarized DPDs $F_{\Delta d \Delta u_1}(x_d,x_{u_1},\vec{y}_\perp)$ as functions of $x_{u_1}$ and $
|\vec{y}_\perp|$ for different values of $x_d=\{0.5,\,4.5,\,7.5,\,11.5\}/16.5$.}
\label{fig9}
\end{figure}

\begin{figure}[tph]
\centering
\includegraphics[width=0.5\textwidth]{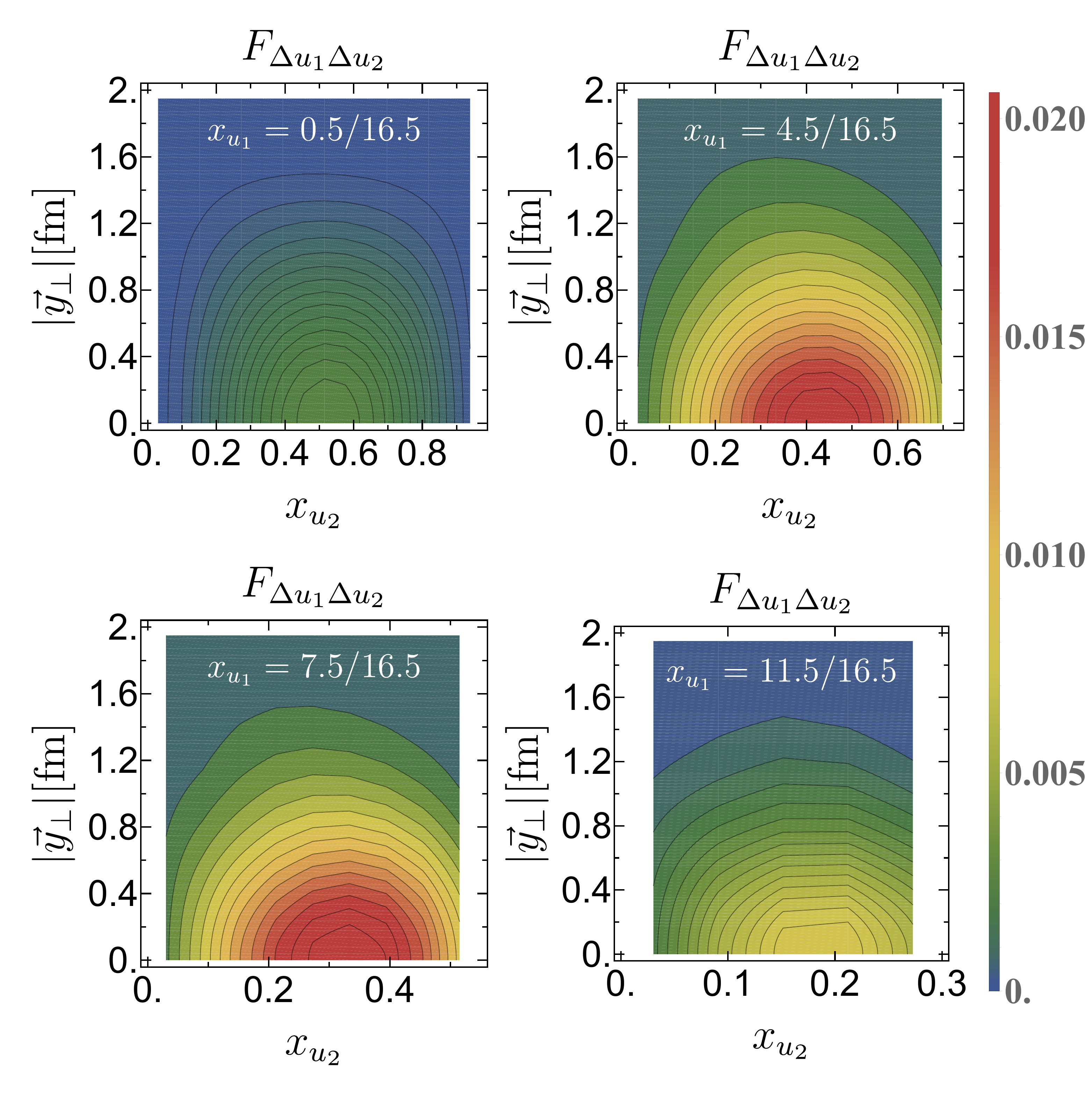}
\caption{The polarized DPDs $F_{\Delta u_1 \Delta u_2}(x_{u_1},x_{u_2},\vec{y}_\perp)$ as functions of $x_{u_2}$ and $
|\vec{y}_\perp|$ for different value of $x_{u_1}=\{0.5,\,4.5,\,7.5,\,11.5\}/16.5$.}
\label{fig10}
\end{figure}

In  Fig.~\ref{fig9}, we show the contour plot of the polarized DPDs $F_{\Delta d \Delta u_1}$ as functions of $x_{u_1}$ and $|\vec{y}_\perp|$ with fixed values of $x_d=\{0.5,\,4.5,\,7.5,\,11.5\}/16.5$.
Similar to the unpolarized DPDs, we observe the maxima of the magnitude of the DPDs when the relative distance between the two quarks vanishes. Figure~\ref{fig10} shows the polarized DPDs $F_{\Delta u_1 \Delta u_2}$ as functions of $x_{u_2}$ and $|\vec{y}_\perp|$ with fixed values of $x_{u_1}$.

\begin{figure}[tph]
\centering
\includegraphics[width=0.45\textwidth]{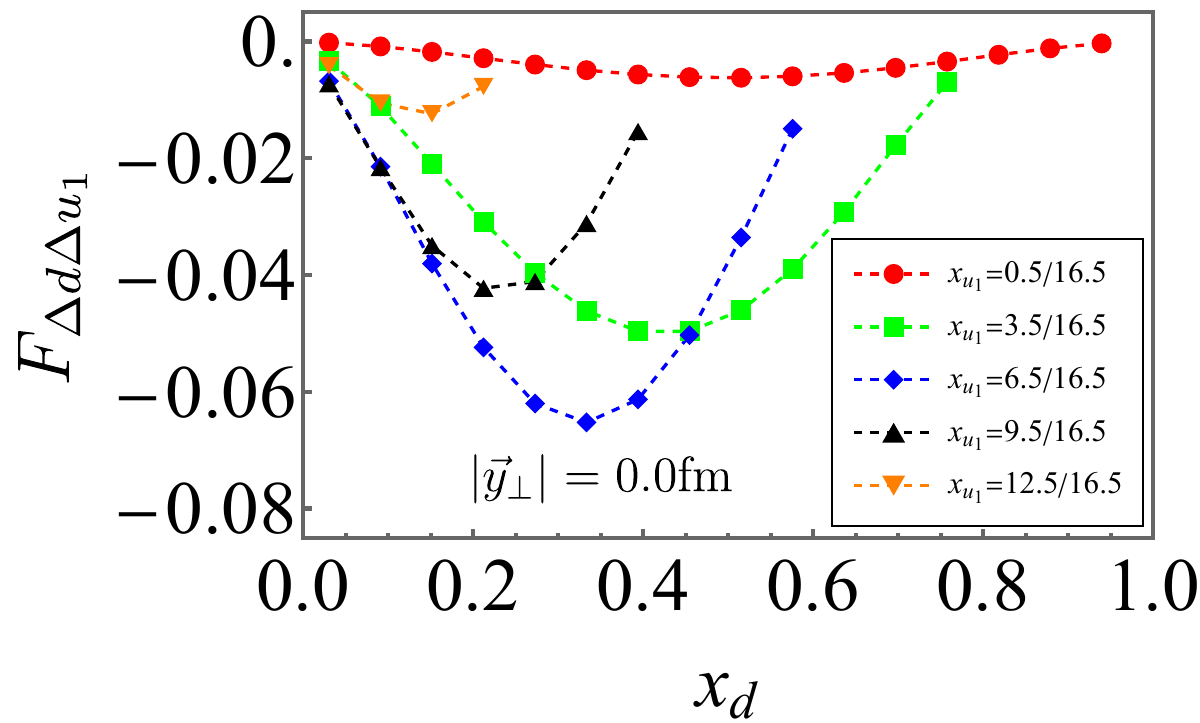}
\includegraphics[width=0.45\textwidth]{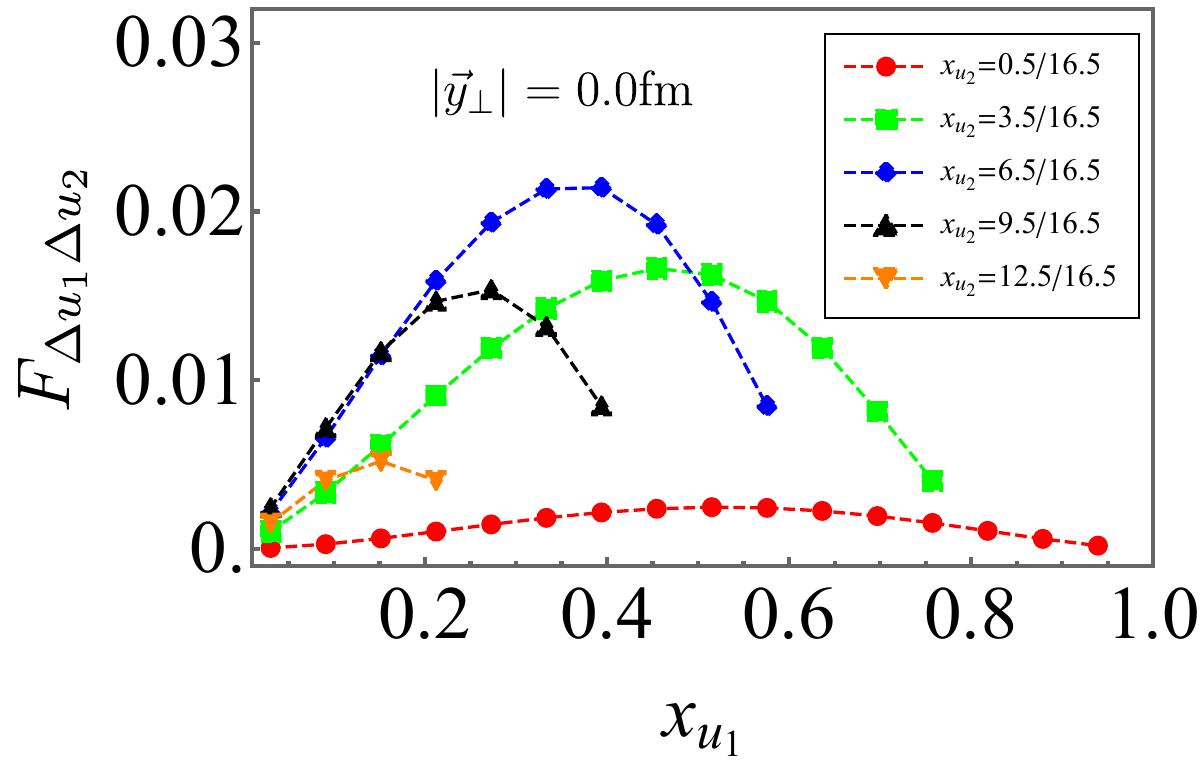}
\caption{The  polarized DPDs,  $F_{\Delta d \Delta u_1}(x_d,x_{u_1},\vec{y}_\perp)$ (upper panel)  and  $F_{\Delta u_1\Delta u_2}(x_{u_1},x_{u_2},\vec{y}_\perp)$ (lower panel), as functions of $x_d (x_{u_1})$ for different values of $x_{u_1} (x_{u_2})=\{0.5,\,3.5,\,6.5,\,9.5,\,12.5\}/16.5$ and fixed $|\vec{y}_\perp|=0$ fm.}
\label{fig11}
\end{figure}

In Fig.~\ref{fig11}, we plot the polarized DPDs as functions of $x_{d}~(x_{u_1})$ for fixed values of $x_{u_1} (x_{u_2})$ at the transverse distance of $|\vec{y}_\perp| = 0 ~\mathrm{fm}$. The qualitative behavior is similar to the unpolarized case as shown in Fig.~\ref{fig5}. As the value of $x_{u_1}(x_{u_2})$ increases, the peak position of the curve shifts towards lower values of $ x_{d}~(x_{u_1})$ and simultaneously the magnitude of the peak increases first and then decreases.

\begin{figure}[tph]
\centering
\includegraphics[width=0.45\textwidth]{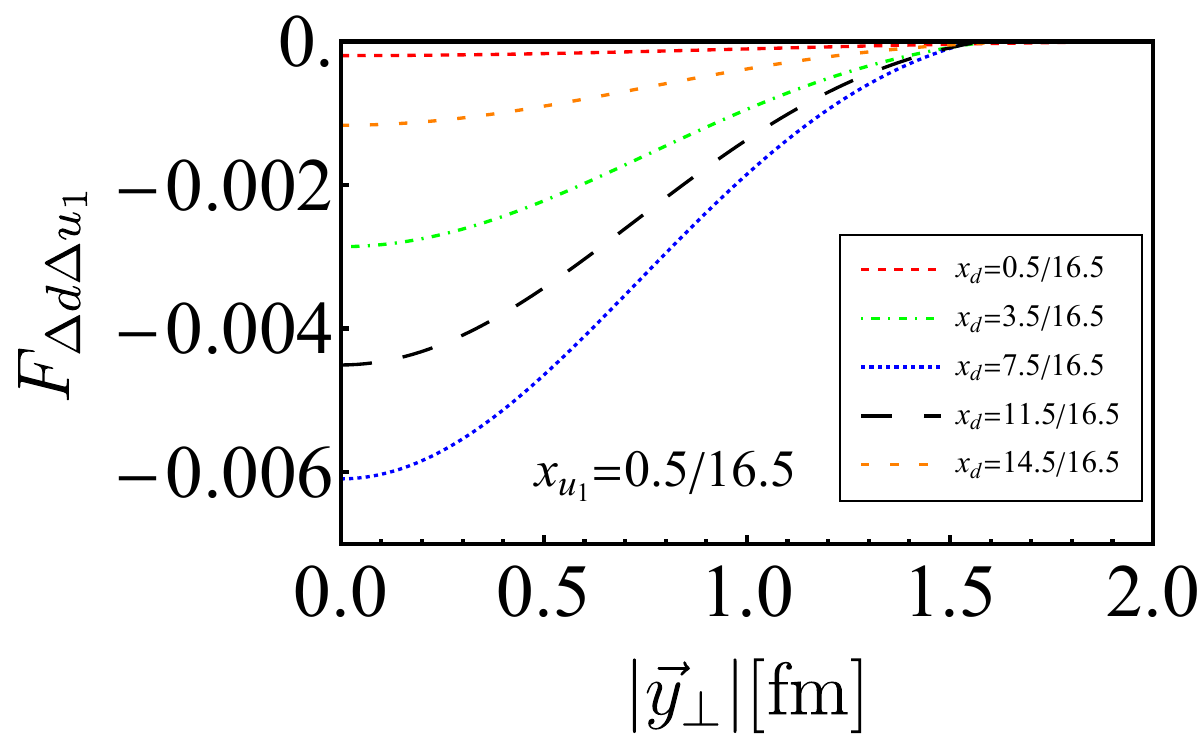}
\includegraphics[width=0.45\textwidth]{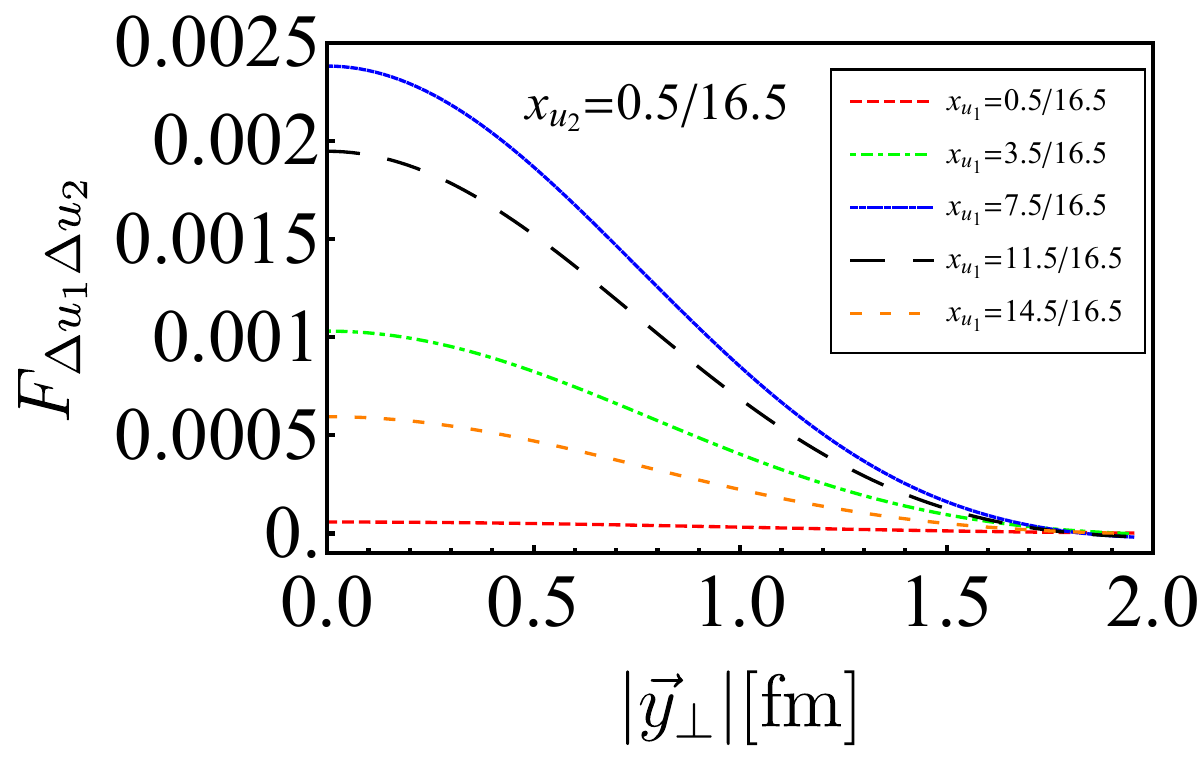}
\caption{The  polarized DPDs,  $F_{\Delta d \Delta u_1}(x_d,x_{u_1},\vec{y}_\perp)$ (upper panel)  and  $F_{\Delta u_1 \Delta u_2}(x_{u_1},x_{u_2},\vec{y}_\perp)$ (lower panel), as functions of $|\vec{y}_\perp|$ for different values of $x_d(x_{u_1})=\{0.5\,,3.5\,,7.5\,,11.5\,,14.5\}/16.5$ and fixed values of $x_{u_1}(x_{u_2})=0.5/16.5$.}
\label{fig12}
\end{figure}

In Fig.~\ref{fig12}, we show the polarized DPDs as functions of $|\vec{y}_\perp|$, while keeping $x_{u_1}$($x_{u_2}$) fixed and varying the values of $x_{d}$ ($x_{u_1}$) for different curves. Our analysis reveals that $|F_{\Delta d \Delta u_1}|$ exceeds $|F_{\Delta u_1 \Delta u_2}|$, indicating a stronger correlation between longitudinally polarized down and up quarks compared to that between longitudinally polarized up and up quarks within the proton.

\subsection{DPDs in momentum space}

The Fourier transform allows the conversion of DPDs from coordinate space to momentum space, as represented by the following equation:

\begin{align}
\int \frac{d^2\vec{y}_\perp}{(2\pi)^2}
e^{-i\vec{k}_\perp \cdot \vec{y}_\perp} F(x_1,x_2,\vec{y}_\perp)=\widetilde{F}(x_1,x_2,\vec{k}_\perp),
\end{align}
where $\widetilde{F}(x_1,x_2,\vec{k}_\perp)$ denotes the DPDs in momentum space. The variable $ \vec{k}_\perp$ represents the difference in transverse momenta between the struck quarks in the initial and final states, expressed as $\vec{k}_\perp=\vec{k_1^\prime}_\perp-\vec{k_1}_\perp=\vec{k_2}_\perp -\vec{k_2^\prime}_\perp$.

\begin{figure}[tph]
\centering
\includegraphics[width=0.45\textwidth]{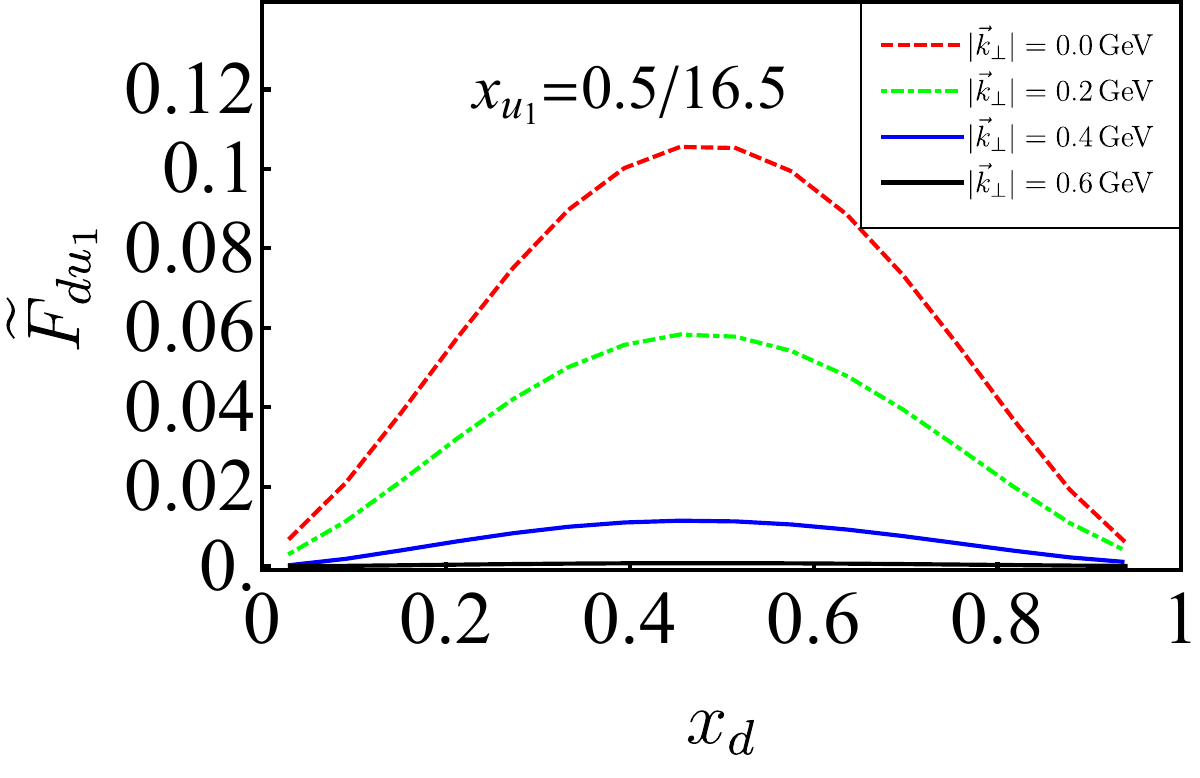}
\includegraphics[width=0.45\textwidth]{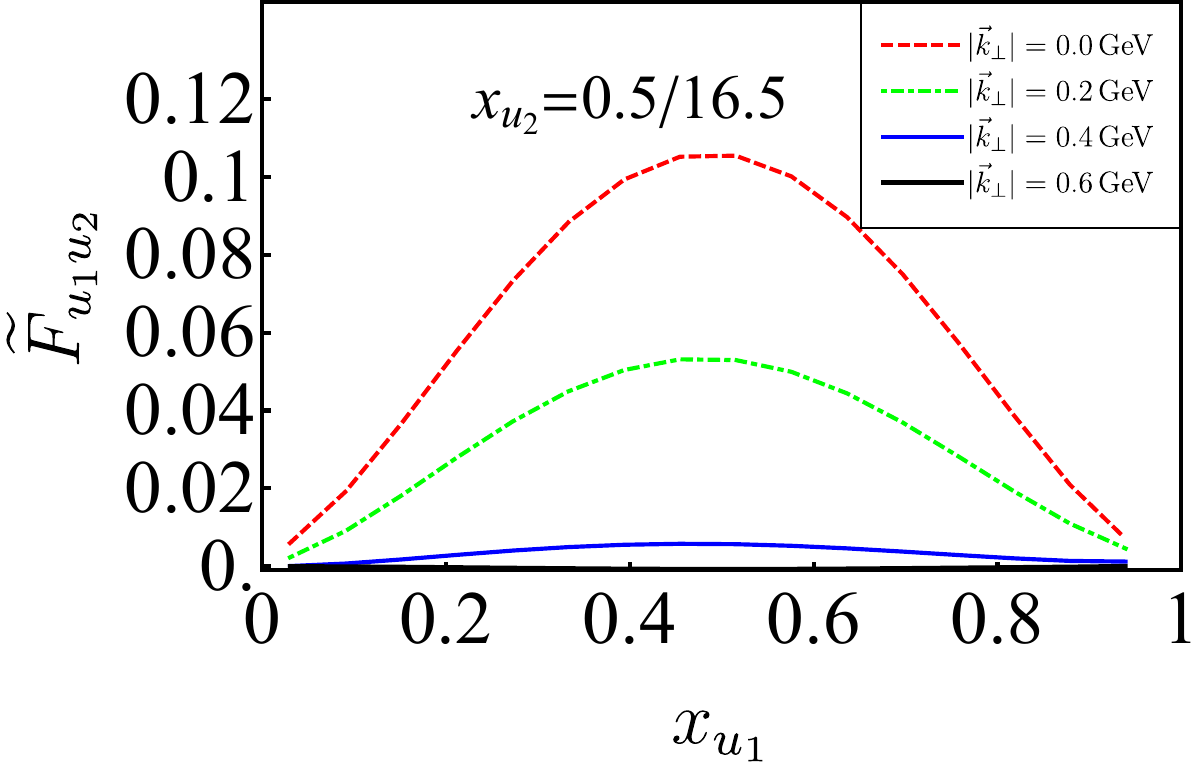}
\caption{The unpolarized DPDs, $\widetilde{F}_{du_1}(x_d,x_{u_1},|\vec{k}_\perp|)$ (upper panel) and $\widetilde{F}_{u_1u_2}(x_{u_1},x_{u_2},|\vec{k}_\perp|)$ (lower panel) as functions of $x_{d}~(x_{u_1})$ for different values of $|\vec{k}_\perp|=\{0,\,0.2,\,0.4,\,0.6\}$ GeV and fixed $x_{u_1}(x_{u_2})=0.5/16.5$.}
\label{fig13}
\end{figure}

\begin{figure}[tph]
\centering
\includegraphics[width=0.45\textwidth]{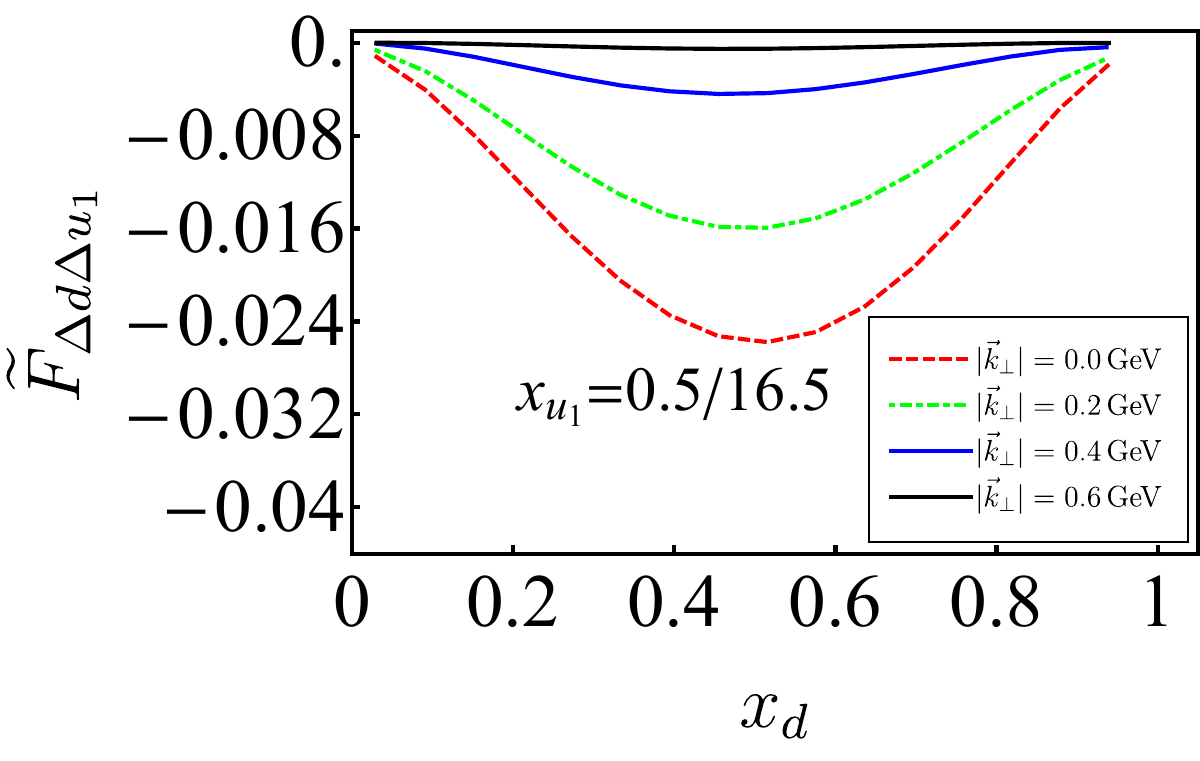}
\includegraphics[width=0.45\textwidth]{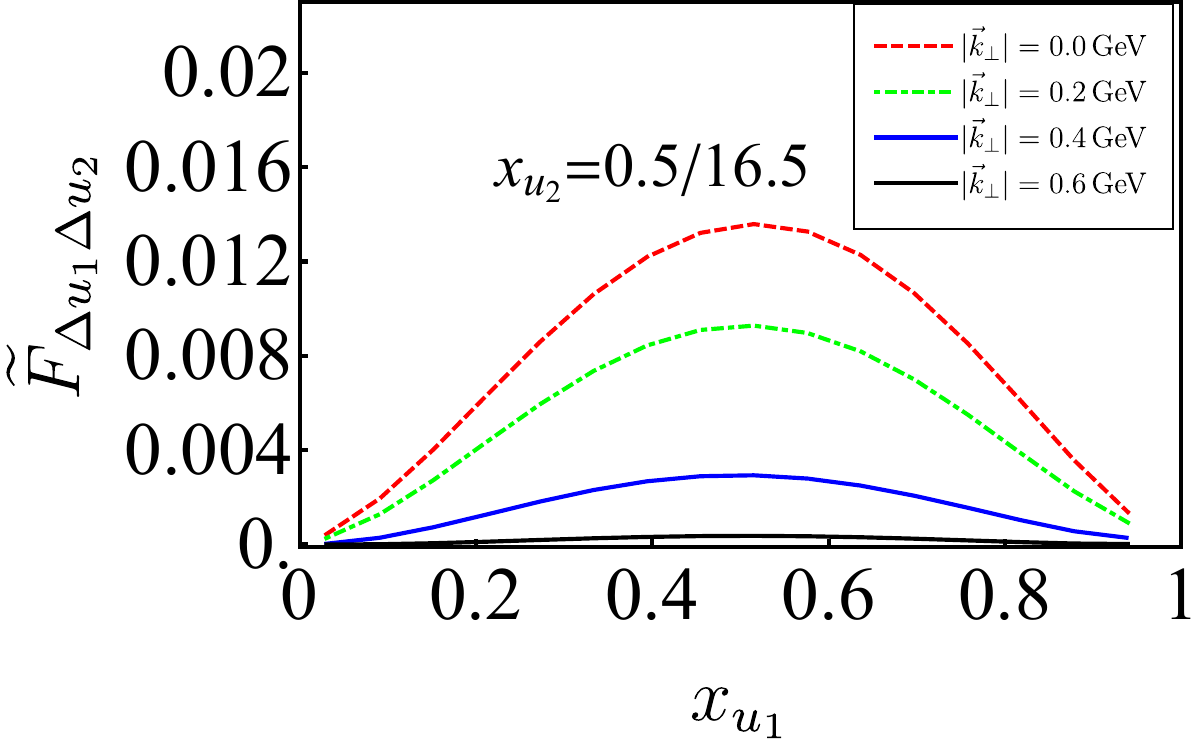}
\caption{The polarized DPDs, $\widetilde{F}_{\Delta d \Delta u_1}(x_d,x_{u_1},|\vec{k}_\perp|)$ (upper panel) and $\widetilde{F}_{\Delta u_1 \Delta u_2}(x_{u_1},x_{u_2},|\vec{k}_\perp|)$ (lower panel) as functions of $x_{d}~(x_{u_1})$ for different values of $|\vec{k}_\perp|=\{0,\,0.2,\,0.4,\,0.6\}$ GeV and fixed $x_{u_1}(x_{u_2})=0.5/16.5$.}
\label{fig14}
\end{figure}

In Fig.~\ref{fig13}, we present the unpolarized DPDs in momentum space, fixing $x_{u_1} (x_{u_2})$ and plotting them as functions of $x_{d}~(x_{u_1})$ for various values of $|\vec{k}_\perp|$. 
As $|\vec{k}_\perp|$ increases, we notice that the peak of the distribution not only gets smaller but also moves slightly towards lower values of $x$. 
At $|\vec{k}_\perp| = 0$, the peak values of $\widetilde{F}_{du_1}$ and $\widetilde{F}_{u_1u_2}$ are almost identical.
However, for non-zero $|\vec{k}_\perp|$, the peak of $\widetilde{F}_{du_1}$ is slightly higher than that of $\widetilde{F}_{u_1u_2}$.
The decrease of the peak value is significant, dropping from its initial value $(0.10553)$ to a much lower $(0.00074)$ one as $|\vec{k}_\perp|$ increases from 0 to 0.6 GeV for $\widetilde{F}_{du_1}$.
The general trends of our findings are consistent with predictions from the bag model~\cite{Chang:2012nw} and the constituent quark model~\cite{Rinaldi:2013vpa, Rinaldi:2014ddl}. Similarly, the polarized DPDs depicted in Fig.~\ref{fig14} exhibit trends akin to those in Fig.~\ref{fig13}, with $\widetilde{F}_{\Delta d \Delta u_1}$ having negative values, in contrast to $\widetilde{F}_{\Delta u_1 \Delta u_2}$.

We further analyze the $|\vec{k}_\perp|$ dependence by evaluating the ratio $\widetilde{R}$ defined as,
\begin{equation}\label{dpd_ratio}	
\begin{split}
\widetilde{R}_{q_1q_2} (x_1 , k_\perp)=&\frac{\widetilde{F}_{q_1q_2}(x_1,x_2=0.4,k_\perp)}{\widetilde{F}_{q_1q_2}(x_1=0.4,x_2=0.4,k_\perp)},\\
\widetilde{R}_{\Delta{q_1}\Delta{q_2}}(x_1 , k_\perp)=&\frac{\widetilde{F}_{\Delta{q_1}\Delta{q_2}}(x_1,x_2=0.4,k_\perp)}{\widetilde{F}_{\Delta{q_1}\Delta{q_2}}(x_1=0.4,x_2=0.4,k_\perp)}.
\end{split}
\end{equation}

This ratio was studied in the bag model~\cite{Chang:2012nw} and the constituent quark model~\cite{Rinaldi:2013vpa, Rinaldi:2014ddl} to test the violation of the factorization of the DPDs into functions of the transverse separation and functions of the longitudinal momentum fractions. This ratio should be independent of $|\vec{k}_\perp|$ if the factorization holds exactly and thus, the magnitude of $|\vec{k}_\perp|$ dependence of this ratio would be a measure of the violation of the factorization in $x$ and $\vec{k}_\perp$.

\begin{figure}[tph]
\centering
\includegraphics[width=0.45\textwidth]{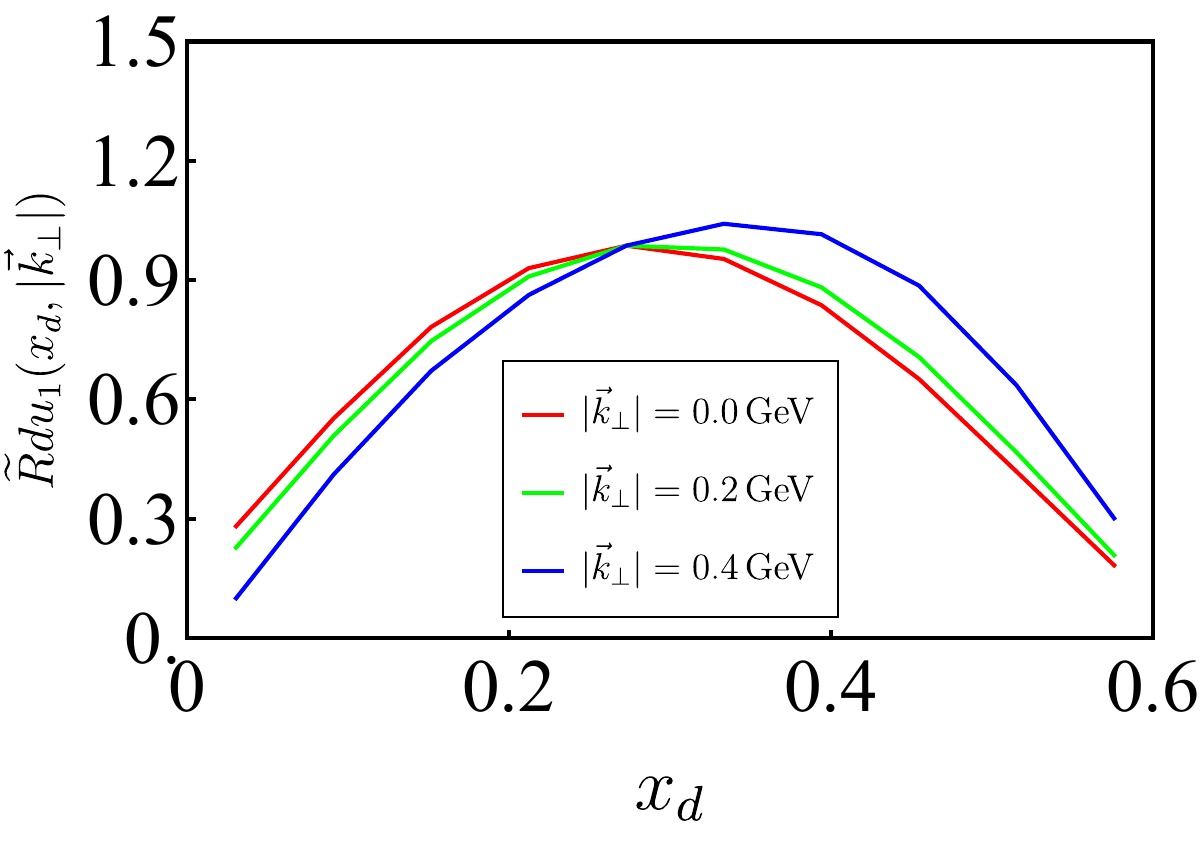}
\includegraphics[width=0.45\textwidth]{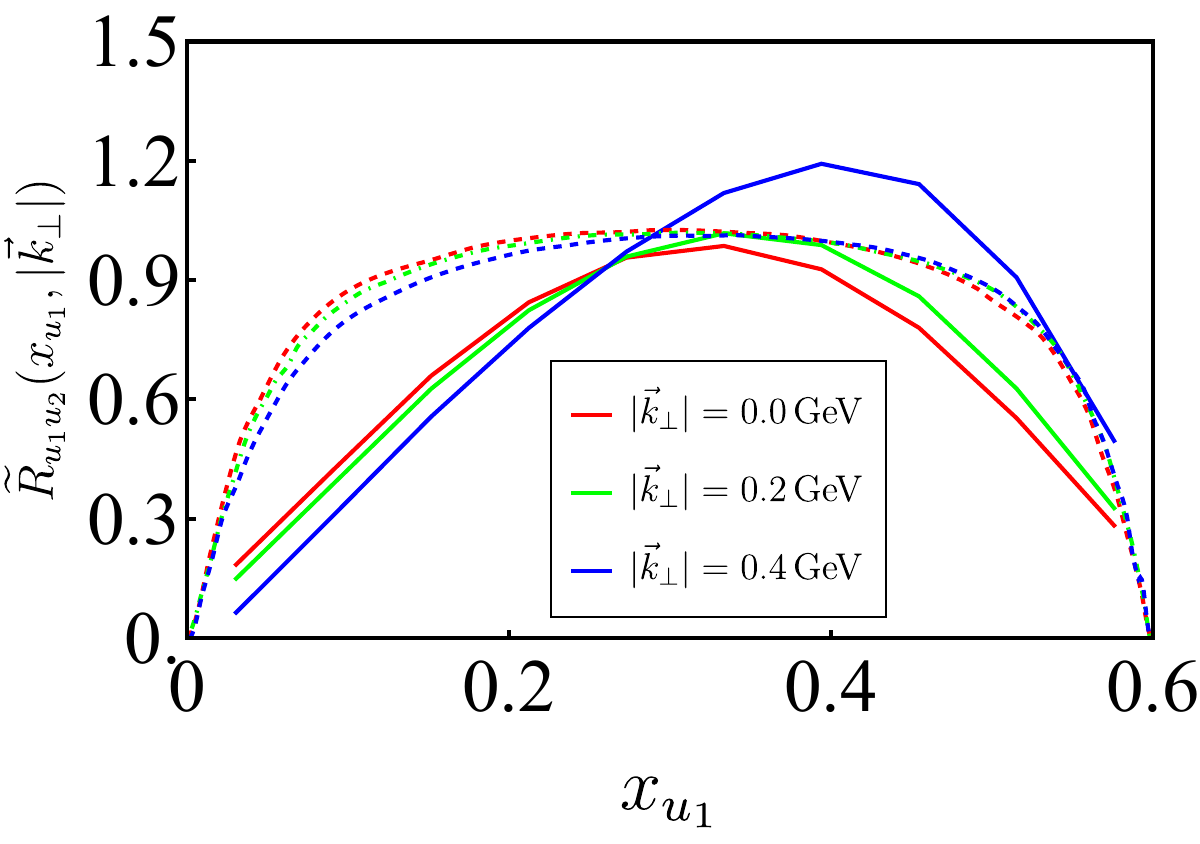}
\caption{The unpolarized DPDs ratio $\widetilde{R}_{du_1}$ (upper panel) and $\widetilde{R}_{u_1u_2}$ (lower panel) as defined in Eq.~\eqref{dpd_ratio} are plotted  as functions of $x_d$($x_{u_1}$), for three values of $|\vec{k}_\perp|=\{0,\,0.2,\,0.4\}$ GeV. Our results (solid lines) for $\widetilde{R}_{u_1u_2}$ are compared with those from the constituent quark model (dashed lines)~\cite{Rinaldi:2014ddl}. }
\label{fig15}
\end{figure}

\begin{figure}[tph]
\centering
\includegraphics[width=0.45\textwidth]{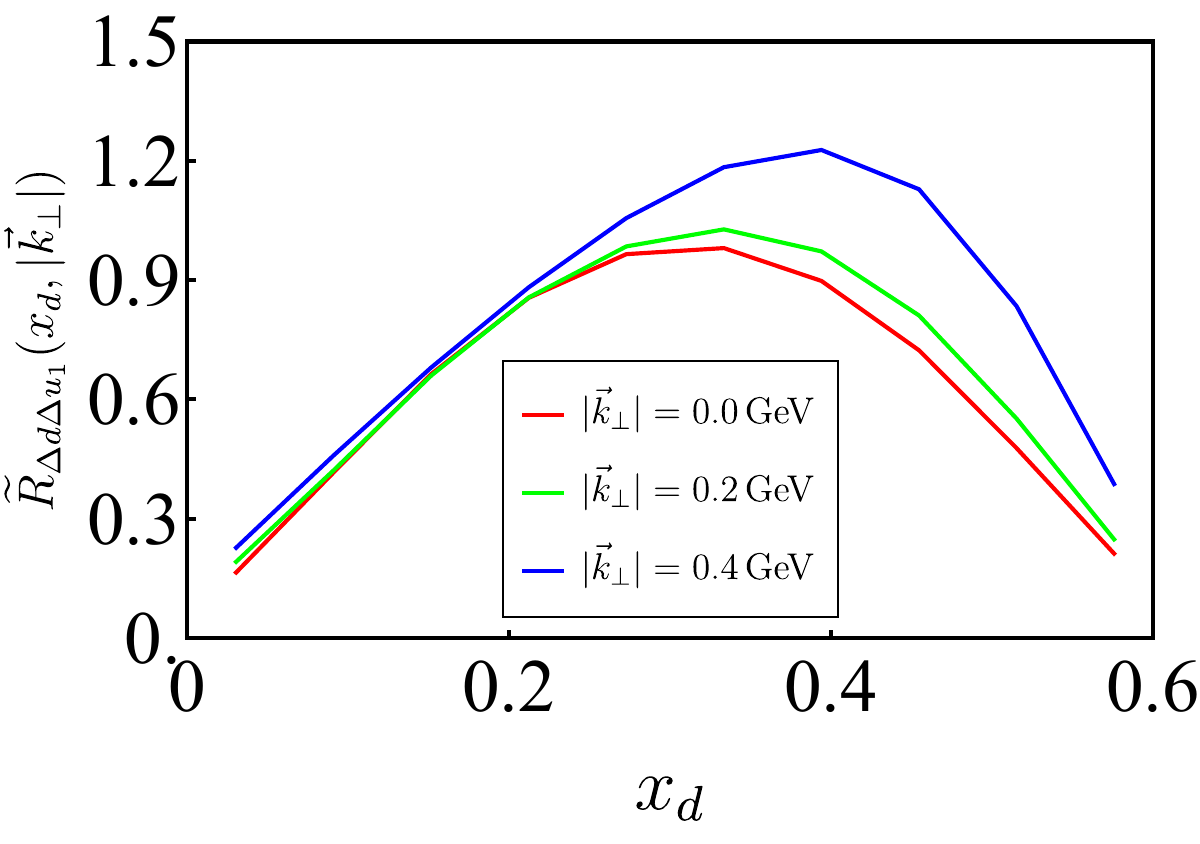}
\includegraphics[width=0.45\textwidth]{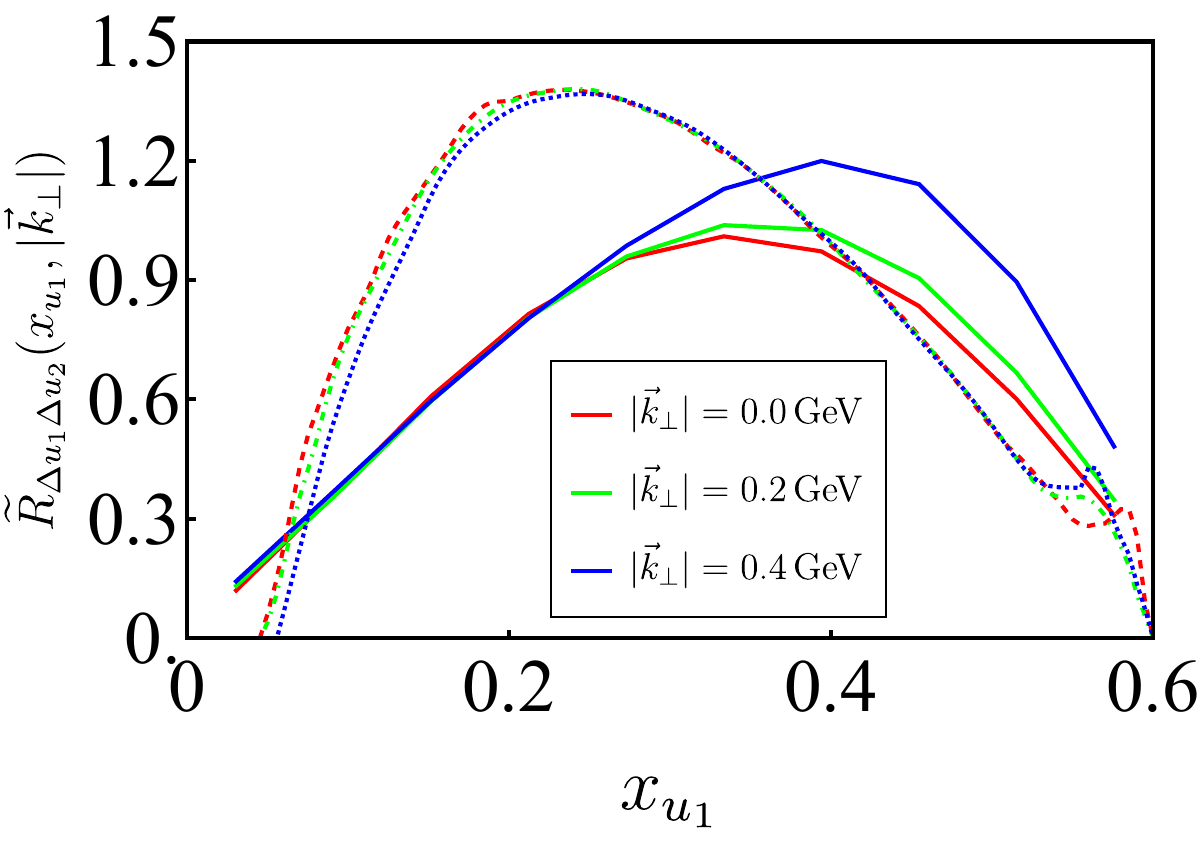}
\caption{The polarized DPDs ratio $\widetilde{R}_{\Delta d \Delta u_1}$ (upper panel) and $\widetilde{R}_{\Delta u_1 \Delta u_2}$ (lower panel) as defined in Eq.~\eqref{dpd_ratio} are plotted  as functions of $x_d$($x_{u_1}$), for three values of $|\vec{k}_\perp|=\{0,\,0.2,\,0.4\}$ GeV. Our results (solid lines) for $\widetilde{R}_{\Delta u_1 \Delta u_2}$ are compared with those from the constituent quark model (dashed lines)~\cite{Rinaldi:2014ddl}.}
\label{fig16}
\end{figure}

In Figs.~\ref{fig15} and \ref{fig16}, we present the unpolarized and polarized ratios of the DPDs, respectively, plotted as functions of the longitudinal momentum fraction at different values of $|\vec{k}_\perp|$. To facilitate comparison with the ratios $\widetilde{R}_{u_1u_2}$ and $\widetilde{R}_{\Delta u_1 \Delta u_2}$ calculated in the constituent quark model~\cite{Rinaldi:2014ddl}, we use 0.4 as the constant parameter in the definition of these ratios, as with Eqs. (3.9, 3.10) in Ref~\cite{Rinaldi:2014ddl}. The most notable observation from these plots of the results from our approach is the significant increase in the factorization violation as $|\vec{k}_\perp|$ increases, particularly evident in the high $x_{u_1}$ region, which is more pronounced in the polarized distributions, shown in Fig.~\ref{fig16}. Both in the constituent quark model~\cite{Rinaldi:2014ddl} and in our results the violation of factorization increases as $|\vec{k}_\perp|$ increases, though the violation is more pronounced with our results. When compared with the ratios obtained in the constituent quark model from Ref.~\cite{Rinaldi:2014ddl}, our results show a shift towards larger $x_{u_1}$ values for both the unpolarized and polarized distributions. 
Furthermore, in the lower $x_{u_1}$ range, the unpolarized ratios exhibit significantly greater factorization violations compared to the polarized ones. This is evident from the larger separations among the lines for $\widetilde{R}_{u_1 u_2}$ in Fig.~\ref{fig15}, in contrast to the smaller separations among the lines for $\widetilde{R}_{\Delta u_1 \Delta u_2}$ in Fig.~\ref{fig16}. In fact, the polarized ratio, especially $\widetilde{R}_{\Delta u_1 \Delta u_2}$, remains nearly constant across different values of $\vec{k}_\perp$ in the low $x_{u_1}$ region.

\section{Summary and outlooks}\label{sec:conclusion}
Basis light-front quantization (BLFQ) is a fully relativistic and nonperturbative method for solving quantum field theory based on the light-front Hamiltonian. Using a recently proposed effective light-front Hamiltonian for the baryon within BLFQ, we presented a study of double parton distributions (DPDs) of the proton. With quarks being the only degrees of freedom, the effective Hamiltonian incorporates confinement in both the transverse and longitudinal directions and a one-gluon exchange interaction for the constituent valence quarks suitable for low-resolution properties. We obtained the light-front wave functions (LFWFs) as the eigenvectors of this effective Hamiltonian in the leading Fock sector. Employing the resulting LFWFs, we calculated the unpolarized and longitudinally polarized quark-quark double parton distributions of the proton. 

Our approach enables a direct investigation into the correlation between quark momentum fractions $x_1$ and $x_2$, and their transverse separation $\vec{y}_\perp$, without assuming any factorization between them. Our results reveal strong correlations among these variables and are promising for enhancing model parametrizations of baryon DPDs.

All unpolarized DPDs are positive, while those involving a down struck quark are negative for polarized DPDs. We observed that the correlation is stronger between two unpolarized up quarks compared to an up and a down quark ($F_{ d u_1} < F_{u_1 u_2}$). Conversely, in the polarized case, the correlation is stronger between polarized down and up quarks than between two polarized up quarks within the proton ($|F_{\Delta d \Delta u_1}| > |F_{\Delta u_1 \Delta u_2}|$).
Furthermore, we presented the DPDs in momentum space, finding consistency with the constituent quark model broadly. Our calculations do not support the $x-\vec{k}_\perp$ factorization of the DPDs, which is often employed in various phenomenological applications and experimental programs, including those at the Large Hadron Collider. 

This study aims to explore the double parton structure of the proton within its valence Fock sector. Future investigations will include higher Fock states, aiming to uncover correlations among valence quarks, sea quarks, and gluons within the proton. The accumulation of high-precision data on double parton scattering from hadron-hadron collision experiments will provide valuable insights into a detailed picture of the multi-parton structure of the proton.  By employing QCD evolution equations for DPDs, we also intend to evolve these distributions to higher energy scales for comparison with double parton scattering cross-sections observed in hadron-hadron collision experiments.   Upcoming research will also examine double parton correlations in heavy baryon systems, exploring potential distinctions from the proton.

\begin{acknowledgments}
S. N. is supported by the Senior Scientist Program funded by Gansu Province, Grant No. 23JDKA0005. C. M. thanks the Chinese Academy of Sciences President’s International Fellowship Initiative for the support via Grants No. 2021PM0023. C. M. is also supported by new faculty start up funding by the Institute of Modern Physics, Chinese Academy of Sciences, Grant No. E129952YR0. 
X.L. is supported by the National Natural Science Foundation of China under Grants No.~12335001 and No.~12247101, National Key Research and Development Program of China under Contract No.~2020YFA0406400, the 111 Project under Grant No.~B20063, the fundamental Research Funds for the Central Universities, and the project for top-notch innovative talents of Gansu province.
X. Z. is supported by new faculty startup funding by the Institute of Modern Physics, Chinese Academy of Sciences, by Key Research Program of Frontier Sciences, Chinese Academy of Sciences, Grant No. ZDBS-LY-7020, by the Foundation for Key Talents of Gansu Province, by the Central Funds Guiding the Local Science and Technology Development of Gansu Province, Grant No. 22ZY1QA006, by international partnership program of the Chinese Academy of Sciences, Grant No. 016GJHZ2022103FN, by National Key R\&D Program of China, Grant No. 2023YFA1606903, by the Strategic Priority Research Program of the Chinese Academy of Sciences, Grant No. XDB34000000, Grant No. 25RCKA008, by the Senior Scientist Program funded by Gansu Province, and by the National Natural Science Foundation of China under Grant No.12375143. 
J. P. V. is supported by the Department of Energy under Grant No. DE-SC0023692. 
This research is supported by Gansu International Collaboration and Talents Recruitment Base of Particle Physics (2023–2027). A major portion of the computational resources were also provided by Sugon Advanced Computing Center.
T. P. is supported by Education Department of Gansu Province for Excellent Graduate Student “Innovation Star” under Grant No. 2025CXZX-043.
This research is supported by Gansu International Collaboration and Talents Recruitment Base of Particle Physics (2023–2027). A major portion of the computational resources were also provided by Sugon Advanced Computing Center.
\end{acknowledgments}


\bibliography{references.bib}

\begin{thebibliography}{100}
\expandafter\ifx\csname url\endcsname\relax
  \def\url#1{\texttt{#1}}\fi
\expandafter\ifx\csname urlprefix\endcsname\relax\def\urlprefix{URL }\fi
\expandafter\ifx\csname href\endcsname\relax
  \def\href#1#2{#2} \def\path#1{#1}\fi

\bibitem{Soper:1996sn}
D.~E. Soper, {Parton distribution functions}, Nucl. Phys. B Proc. Suppl. 53
  (1997) 69--80.
\newblock \href {http://arxiv.org/abs/hep-lat/9609018}
  {\path{arXiv:hep-lat/9609018}}, \href
  {https://doi.org/10.1016/S0920-5632(96)00600-7}
  {\path{doi:10.1016/S0920-5632(96)00600-7}}.

\bibitem{Bartalini:2018qje}
P.~Bartalini, J.~R. Gaunt (Eds.), {Multiple Parton Interactions at the LHC},
  Vol.~29, WSP, 2019.
\newblock \href {https://doi.org/10.1142/10646} {\path{doi:10.1142/10646}}.

\bibitem{CMS:2022pio}
A.~Tumasyan, et~al., {Observation of same-sign WW production from double parton
  scattering in proton-proton collisions at $\sqrt{s}$ = 13 TeV}, Phys. Rev.
  Lett. 131 (2023) 091803.
\newblock \href {http://arxiv.org/abs/2206.02681} {\path{arXiv:2206.02681}},
  \href {https://doi.org/10.1103/PhysRevLett.131.091803}
  {\path{doi:10.1103/PhysRevLett.131.091803}}.

\bibitem{Leontsinis:2022cyi}
S.~Leontsinis, {Observation of triple J/\ensuremath{\psi} meson production in
  proton-proton collisions at $\sqrt {s}$=13 TeV}, Rev. Mex. Fis. Suppl. 3~(3)
  (2022) 0308046.
\newblock \href {https://doi.org/10.31349/SuplRevMexFis.3.0308046}
  {\path{doi:10.31349/SuplRevMexFis.3.0308046}}.

\bibitem{CMS:2021lxi}
A.~Tumasyan, et~al., {Measurement of double-parton scattering in inclusive
  production of four jets with low transverse momentum in proton-proton
  collisions at $ \sqrt{s} $ = 13 TeV}, JHEP 01 (2022) 177.
\newblock \href {http://arxiv.org/abs/2109.13822} {\path{arXiv:2109.13822}},
  \href {https://doi.org/10.1007/JHEP01(2022)177}
  {\path{doi:10.1007/JHEP01(2022)177}}.

\bibitem{LHCb:2020jse}
R.~Aaij, et~al., {Observation of Enhanced Double Parton Scattering in
  Proton-Lead Collisions at $\sqrt {s_{NN}}$ =8.16 TeV}, Phys. Rev. Lett.
  125~(21) (2020) 212001.
\newblock \href {http://arxiv.org/abs/2007.06945} {\path{arXiv:2007.06945}},
  \href {https://doi.org/10.1103/PhysRevLett.125.212001}
  {\path{doi:10.1103/PhysRevLett.125.212001}}.

\bibitem{CMS:2019jcb}
A.~M. Sirunyan, et~al., {Evidence for $\text {W}\text {W}$ production from
  double-parton interactions in proton\textendash{}proton collisions at
  $\sqrt{s} = 13 \,\text {TeV} $}, Eur. Phys. J. C 80~(1) (2020) 41.
\newblock \href {http://arxiv.org/abs/1909.06265} {\path{arXiv:1909.06265}},
  \href {https://doi.org/10.1140/epjc/s10052-019-7541-6}
  {\path{doi:10.1140/epjc/s10052-019-7541-6}}.

\bibitem{Belyaev:2017sws}
I.~Belyaev, D.~Savrina, {Study of double parton scattering processes with heavy
  quarks}, Adv. Ser. Direct. High Energy Phys. 29 (2018) 141--157.
\newblock \href {http://arxiv.org/abs/1711.10877} {\path{arXiv:1711.10877}},
  \href {https://doi.org/10.1142/9789813227767_0008}
  {\path{doi:10.1142/9789813227767_0008}}.

\bibitem{CMS:2017han}
A.~M. Sirunyan, et~al., {Constraints on the double-parton scattering cross
  section from same-sign W boson pair production in proton-proton collisions at
  $ \sqrt{s}=8 $ TeV}, JHEP 02 (2018) 032.
\newblock \href {http://arxiv.org/abs/1712.02280} {\path{arXiv:1712.02280}},
  \href {https://doi.org/10.1007/JHEP02(2018)032}
  {\path{doi:10.1007/JHEP02(2018)032}}.

\bibitem{An:2017kyn}
L.~An, {Measurements of soft QCD and double parton scattering at LHCb}, in:
  {5th Large Hadron Collider Physics Conference}, 2017.
\newblock \href {http://arxiv.org/abs/1708.09773} {\path{arXiv:1708.09773}}.

\bibitem{Koshkarev:2022mgi}
S.~Koshkarev, S.~Groote, {Double J/\ensuremath{\psi} production as a test of
  parton momentum correlations in double parton scattering}, Phys. Lett. B 847
  (2023) 138296.
\newblock \href {http://arxiv.org/abs/2208.13429} {\path{arXiv:2208.13429}},
  \href {https://doi.org/10.1016/j.physletb.2023.138296}
  {\path{doi:10.1016/j.physletb.2023.138296}}.

\bibitem{Kirschner:1979im}
R.~Kirschner, {Generalized {Lipatov-Altarelli-Parisi} Equations and Jet
  Calculus Rules}, Phys. Lett. B 84 (1979) 266--270.
\newblock \href {https://doi.org/10.1016/0370-2693(79)90300-9}
  {\path{doi:10.1016/0370-2693(79)90300-9}}.

\bibitem{SHELEST1982325}
V.~Shelest, A.~Snigirev, G.~Zinovjev,
  \href{https://www.sciencedirect.com/science/article/pii/0370269382900491}{Gazing
  into the multiparton distribution equations in qcd}, Physics Letters B
  113~(4) (1982) 325--328.
\newblock \href {https://doi.org/https://doi.org/10.1016/0370-2693(82)90049-1}
  {\path{doi:https://doi.org/10.1016/0370-2693(82)90049-1}}.
\newline\urlprefix\url{https://www.sciencedirect.com/science/article/pii/0370269382900491}

\bibitem{Zinovev:1982be}
G.~m. Zinovev, A.~m. Snigirev, V.~p. Shelest, {EQUATIONS FOR MANY PARTON
  DISTRIBUTIONS IN QUANTUM CHROMODYNAMICS}, Theor. Math. Phys. 51 (1982)
  523--528.
\newblock \href {https://doi.org/10.1007/BF01017270}
  {\path{doi:10.1007/BF01017270}}.

\bibitem{Snigirev:2003cq}
A.~M. Snigirev, {Double parton distributions in the leading logarithm
  approximation of perturbative QCD}, Phys. Rev. D 68 (2003) 114012.
\newblock \href {http://arxiv.org/abs/hep-ph/0304172}
  {\path{arXiv:hep-ph/0304172}}, \href
  {https://doi.org/10.1103/PhysRevD.68.114012}
  {\path{doi:10.1103/PhysRevD.68.114012}}.

\bibitem{Korotkikh:2004bz}
V.~L. Korotkikh, A.~M. Snigirev, {Double parton correlations versus factorized
  distributions}, Phys. Lett. B 594 (2004) 171--176.
\newblock \href {http://arxiv.org/abs/hep-ph/0404155}
  {\path{arXiv:hep-ph/0404155}}, \href
  {https://doi.org/10.1016/j.physletb.2004.05.012}
  {\path{doi:10.1016/j.physletb.2004.05.012}}.

\bibitem{Ceccopieri:2010kg}
F.~A. Ceccopieri, {An update on the evolution of double parton distributions},
  Phys. Lett. B 697 (2011) 482--487.
\newblock \href {http://arxiv.org/abs/1011.6586} {\path{arXiv:1011.6586}},
  \href {https://doi.org/10.1016/j.physletb.2011.02.047}
  {\path{doi:10.1016/j.physletb.2011.02.047}}.

\bibitem{Ceccopieri:2014ufa}
F.~A. Ceccopieri, {A second update on double parton distributions}, Phys. Lett.
  B 734 (2014) 79--85.
\newblock \href {http://arxiv.org/abs/1403.2167} {\path{arXiv:1403.2167}},
  \href {https://doi.org/10.1016/j.physletb.2014.05.015}
  {\path{doi:10.1016/j.physletb.2014.05.015}}.

\bibitem{Mekhfi:1985dv}
M.~Mekhfi, {Correlations in Color and Spin in Multiparton Processes}, Phys.
  Rev. D 32 (1985) 2380.
\newblock \href {https://doi.org/10.1103/PhysRevD.32.2380}
  {\path{doi:10.1103/PhysRevD.32.2380}}.

\bibitem{Mekhfi:1988kj}
M.~Mekhfi, X.~Artru, {Sudakov Suppression of Color Correlations in Multiparton
  Scattering}, Phys. Rev. D 37 (1988) 2618--2622.
\newblock \href {https://doi.org/10.1103/PhysRevD.37.2618}
  {\path{doi:10.1103/PhysRevD.37.2618}}.

\bibitem{Manohar:2012jr}
A.~V. Manohar, W.~J. Waalewijn, {A QCD Analysis of Double Parton Scattering:
  Color Correlations, Interference Effects and Evolution}, Phys. Rev. D 85
  (2012) 114009.
\newblock \href {http://arxiv.org/abs/1202.3794} {\path{arXiv:1202.3794}},
  \href {https://doi.org/10.1103/PhysRevD.85.114009}
  {\path{doi:10.1103/PhysRevD.85.114009}}.

\bibitem{Diehl:2022rxb}
M.~Diehl, F.~Fabry, A.~Vladimirov, {Two-loop evolution kernels for colour
  dependent double parton distributions}, JHEP 05 (2023) 067.
\newblock \href {http://arxiv.org/abs/2212.11843} {\path{arXiv:2212.11843}},
  \href {https://doi.org/10.1007/JHEP05(2023)067}
  {\path{doi:10.1007/JHEP05(2023)067}}.

\bibitem{CDF:1997yfa}
F.~Abe, et~al., {Double parton scattering in $\bar{p}p$ collisions at $\sqrt{s}
  = 1.8 $TeV}, Phys. Rev. D 56 (1997) 3811--3832.
\newblock \href {https://doi.org/10.1103/PhysRevD.56.3811}
  {\path{doi:10.1103/PhysRevD.56.3811}}.

\bibitem{D0:2015rpo}
V.~M. Abazov, et~al., {Study of double parton interactions in diphoton + dijet
  events in $p\bar{p}$ collisions at $\sqrt{s} = 1.96$ TeV}, Phys. Rev. D
  93~(5) (2016) 052008.
\newblock \href {http://arxiv.org/abs/1512.05291} {\path{arXiv:1512.05291}},
  \href {https://doi.org/10.1103/PhysRevD.93.052008}
  {\path{doi:10.1103/PhysRevD.93.052008}}.

\bibitem{LHCb:2016wuo}
R.~Aaij, et~al., {Measurement of the J/$\psi$ pair production cross-section in
  pp collisions at $ \sqrt{s}=13 $ TeV}, JHEP 06 (2017) 047, [Erratum: JHEP 10,
  068 (2017)].
\newblock \href {http://arxiv.org/abs/1612.07451} {\path{arXiv:1612.07451}},
  \href {https://doi.org/10.1007/JHEP06(2017)047}
  {\path{doi:10.1007/JHEP06(2017)047}}.

\bibitem{ATLAS:2019jzd}
M.~Aaboud, et~al., {Measurement of J/\ensuremath{\psi} production in
  association with a W$^{±}$ boson with pp data at 8 TeV}, JHEP 01 (2020) 095.
\newblock \href {http://arxiv.org/abs/1909.13626} {\path{arXiv:1909.13626}},
  \href {https://doi.org/10.1007/JHEP01(2020)095}
  {\path{doi:10.1007/JHEP01(2020)095}}.

\bibitem{AbdulKhalek:2021gbh}
R.~Abdul~Khalek, et~al., {Science Requirements and Detector Concepts for the
  Electron-Ion Collider}: {EIC Yellow Report}, Nucl. Phys. A 1026 (2022)
  122447.
\newblock \href {http://arxiv.org/abs/2103.05419} {\path{arXiv:2103.05419}},
  \href {https://doi.org/10.1016/j.nuclphysa.2022.122447}
  {\path{doi:10.1016/j.nuclphysa.2022.122447}}.

\bibitem{Anderle:2021wcy}
D.~P. Anderle, et~al., {Electron-ion collider in China}, Front. Phys. (Beijing)
  16~(6) (2021) 64701.
\newblock \href {http://arxiv.org/abs/2102.09222} {\path{arXiv:2102.09222}},
  \href {https://doi.org/10.1007/s11467-021-1062-0}
  {\path{doi:10.1007/s11467-021-1062-0}}.

\bibitem{Katich:2013atq}
J.~Katich, et~al., {Measurement of the Target-Normal Single-Spin Asymmetry in
  Deep-Inelastic Scattering from the Reaction
  $^{3}\mathrm{He}^{\uparrow}(e,e')X$}, Phys. Rev. Lett. 113~(2) (2014) 022502.
\newblock \href {http://arxiv.org/abs/1311.0197} {\path{arXiv:1311.0197}},
  \href {https://doi.org/10.1103/PhysRevLett.113.022502}
  {\path{doi:10.1103/PhysRevLett.113.022502}}.

\bibitem{Metz:2012ui}
A.~Metz, D.~Pitonyak, A.~Schafer, M.~Schlegel, W.~Vogelsang, J.~Zhou,
  {Single-spin asymmetries in inclusive deep inelastic scattering and
  multiparton correlations in the nucleon}, Phys. Rev. D 86 (2012) 094039.
\newblock \href {http://arxiv.org/abs/1209.3138} {\path{arXiv:1209.3138}},
  \href {https://doi.org/10.1103/PhysRevD.86.094039}
  {\path{doi:10.1103/PhysRevD.86.094039}}.

\bibitem{Afanasev:2007ii}
A.~Afanasev, M.~Strikman, C.~Weiss, {Transverse target spin asymmetry in
  inclusive DIS with two-photon exchange}, Phys. Rev. D 77 (2008) 014028.
\newblock \href {http://arxiv.org/abs/0709.0901} {\path{arXiv:0709.0901}},
  \href {https://doi.org/10.1103/PhysRevD.77.014028}
  {\path{doi:10.1103/PhysRevD.77.014028}}.

\bibitem{Paver:1982yp}
N.~Paver, D.~Treleani, {Multi - Quark Scattering and Large $p_T$ Jet Production
  in Hadronic Collisions}, Nuovo Cim. A 70 (1982) 215.
\newblock \href {https://doi.org/10.1007/BF02814035}
  {\path{doi:10.1007/BF02814035}}.

\bibitem{PhysRevD.32.2371}
M.~Mekhfi, \href{https://link.aps.org/doi/10.1103/PhysRevD.32.2371}{Multiparton
  processes: An application to the double drell-yan mechanism}, Phys. Rev. D 32
  (1985) 2371--2379.
\newblock \href {https://doi.org/10.1103/PhysRevD.32.2371}
  {\path{doi:10.1103/PhysRevD.32.2371}}.
\newline\urlprefix\url{https://link.aps.org/doi/10.1103/PhysRevD.32.2371}

\bibitem{SJOSTRAND1987149}
T.~Sjöstrand, M.~{Van Zijl},
  \href{https://www.sciencedirect.com/science/article/pii/0370269387907222}{Multiple
  parton-parton interactions in an impact parameter picture}, Physics Letters B
  188~(1) (1987) 149--154.
\newblock \href {https://doi.org/https://doi.org/10.1016/0370-2693(87)90722-2}
  {\path{doi:https://doi.org/10.1016/0370-2693(87)90722-2}}.
\newline\urlprefix\url{https://www.sciencedirect.com/science/article/pii/0370269387907222}

\bibitem{Blok:2010ge}
B.~Blok, Y.~Dokshitzer, L.~Frankfurt, M.~Strikman, {The Four jet production at
  LHC and Tevatron in QCD}, Phys. Rev. D 83 (2011) 071501.
\newblock \href {http://arxiv.org/abs/1009.2714} {\path{arXiv:1009.2714}},
  \href {https://doi.org/10.1103/PhysRevD.83.071501}
  {\path{doi:10.1103/PhysRevD.83.071501}}.

\bibitem{Gaunt:2011xd}
J.~R. Gaunt, W.~J. Stirling, {Double Parton Scattering Singularity in One-Loop
  Integrals}, JHEP 06 (2011) 048.
\newblock \href {http://arxiv.org/abs/1103.1888} {\path{arXiv:1103.1888}},
  \href {https://doi.org/10.1007/JHEP06(2011)048}
  {\path{doi:10.1007/JHEP06(2011)048}}.

\bibitem{Ryskin:2011kk}
M.~G. Ryskin, A.~M. Snigirev, {A Fresh look at double parton scattering}, Phys.
  Rev. D 83 (2011) 114047.
\newblock \href {http://arxiv.org/abs/1103.3495} {\path{arXiv:1103.3495}},
  \href {https://doi.org/10.1103/PhysRevD.83.114047}
  {\path{doi:10.1103/PhysRevD.83.114047}}.

\bibitem{Blok:2011bu}
B.~Blok, Y.~Dokshitser, L.~Frankfurt, M.~Strikman, {pQCD physics of multiparton
  interactions}, Eur. Phys. J. C 72 (2012) 1963.
\newblock \href {http://arxiv.org/abs/1106.5533} {\path{arXiv:1106.5533}},
  \href {https://doi.org/10.1140/epjc/s10052-012-1963-8}
  {\path{doi:10.1140/epjc/s10052-012-1963-8}}.

\bibitem{Diehl:2011yj}
M.~Diehl, D.~Ostermeier, A.~Schafer, {Elements of a theory for multiparton
  interactions in QCD}, JHEP 03 (2012) 089, [Erratum: JHEP 03, 001 (2016)].
\newblock \href {http://arxiv.org/abs/1111.0910} {\path{arXiv:1111.0910}},
  \href {https://doi.org/10.1007/JHEP03(2012)089}
  {\path{doi:10.1007/JHEP03(2012)089}}.

\bibitem{Manohar:2012pe}
A.~V. Manohar, W.~J. Waalewijn, {What is Double Parton Scattering?}, Phys.
  Lett. B 713 (2012) 196--201.
\newblock \href {http://arxiv.org/abs/1202.5034} {\path{arXiv:1202.5034}},
  \href {https://doi.org/10.1016/j.physletb.2012.05.044}
  {\path{doi:10.1016/j.physletb.2012.05.044}}.

\bibitem{Ryskin:2012qx}
M.~G. Ryskin, A.~M. Snigirev, {Double parton scattering in double logarithm
  approximation of perturbative QCD}, Phys. Rev. D 86 (2012) 014018.
\newblock \href {http://arxiv.org/abs/1203.2330} {\path{arXiv:1203.2330}},
  \href {https://doi.org/10.1103/PhysRevD.86.014018}
  {\path{doi:10.1103/PhysRevD.86.014018}}.

\bibitem{Gaunt:2009re}
J.~R. Gaunt, W.~J. Stirling, {Double Parton Distributions Incorporating
  Perturbative QCD Evolution and Momentum and Quark Number Sum Rules}, JHEP 03
  (2010) 005.
\newblock \href {http://arxiv.org/abs/0910.4347} {\path{arXiv:0910.4347}},
  \href {https://doi.org/10.1007/JHEP03(2010)005}
  {\path{doi:10.1007/JHEP03(2010)005}}.

\bibitem{Golec-Biernat:2014bva}
K.~Golec-Biernat, E.~Lewandowska, {How to impose initial conditions for QCD
  evolution of double parton distributions?}, Phys. Rev. D 90~(1) (2014)
  014032.
\newblock \href {http://arxiv.org/abs/1402.4079} {\path{arXiv:1402.4079}},
  \href {https://doi.org/10.1103/PhysRevD.90.014032}
  {\path{doi:10.1103/PhysRevD.90.014032}}.

\bibitem{Golec-Biernat:2015aza}
K.~Golec-Biernat, E.~Lewandowska, M.~Serino, Z.~Snyder, A.~M. Stasto,
  {Constraining the double gluon distribution by the single gluon
  distribution}, Phys. Lett. B 750 (2015) 559--564.
\newblock \href {http://arxiv.org/abs/1507.08583} {\path{arXiv:1507.08583}},
  \href {https://doi.org/10.1016/j.physletb.2015.09.067}
  {\path{doi:10.1016/j.physletb.2015.09.067}}.

\bibitem{Diehl:2020xyg}
M.~Diehl, J.~R. Gaunt, D.~M. Lang, P.~Pl\"o\ss{}l, A.~Sch\"afer, {Sum rule
  improved double parton distributions in position space}, Eur. Phys. J. C
  80~(5) (2020) 468.
\newblock \href {http://arxiv.org/abs/2001.10428} {\path{arXiv:2001.10428}},
  \href {https://doi.org/10.1140/epjc/s10052-020-8038-z}
  {\path{doi:10.1140/epjc/s10052-020-8038-z}}.

\bibitem{Diehl:2011tt}
M.~Diehl, A.~Schafer, {Theoretical considerations on multiparton interactions
  in QCD}, Phys. Lett. B 698 (2011) 389--402.
\newblock \href {http://arxiv.org/abs/1102.3081} {\path{arXiv:1102.3081}},
  \href {https://doi.org/10.1016/j.physletb.2011.03.024}
  {\path{doi:10.1016/j.physletb.2011.03.024}}.

\bibitem{Diehl:2019rdh}
M.~Diehl, J.~R. Gaunt, P.~Pl\"o\ss{}l, A.~Sch\"afer, {Two-loop splitting in
  double parton distributions}, SciPost Phys. 7~(2) (2019) 017.
\newblock \href {http://arxiv.org/abs/1902.08019} {\path{arXiv:1902.08019}},
  \href {https://doi.org/10.21468/SciPostPhys.7.2.017}
  {\path{doi:10.21468/SciPostPhys.7.2.017}}.

\bibitem{Chang:2012nw}
H.-M. Chang, A.~V. Manohar, W.~J. Waalewijn, {Double Parton Correlations in the
  Bag Model}, Phys. Rev. D 87~(3) (2013) 034009.
\newblock \href {http://arxiv.org/abs/1211.3132} {\path{arXiv:1211.3132}},
  \href {https://doi.org/10.1103/PhysRevD.87.034009}
  {\path{doi:10.1103/PhysRevD.87.034009}}.

\bibitem{Rinaldi:2013vpa}
M.~Rinaldi, S.~Scopetta, V.~Vento, {Double parton correlations in constituent
  quark models}, Phys. Rev. D 87 (2013) 114021.
\newblock \href {http://arxiv.org/abs/1302.6462} {\path{arXiv:1302.6462}},
  \href {https://doi.org/10.1103/PhysRevD.87.114021}
  {\path{doi:10.1103/PhysRevD.87.114021}}.

\bibitem{Broniowski:2013xba}
W.~Broniowski, E.~Ruiz~Arriola, {Valence double parton distributions of the
  nucleon in a simple model}, Few Body Syst. 55 (2014) 381--387.
\newblock \href {http://arxiv.org/abs/1310.8419} {\path{arXiv:1310.8419}},
  \href {https://doi.org/10.1007/s00601-014-0840-4}
  {\path{doi:10.1007/s00601-014-0840-4}}.

\bibitem{Rinaldi:2014ddl}
M.~Rinaldi, S.~Scopetta, M.~Traini, V.~Vento, {Double parton correlations and
  constituent quark models: a Light Front approach to the valence sector}, JHEP
  12 (2014) 028.
\newblock \href {http://arxiv.org/abs/1409.1500} {\path{arXiv:1409.1500}},
  \href {https://doi.org/10.1007/JHEP12(2014)028}
  {\path{doi:10.1007/JHEP12(2014)028}}.

\bibitem{Broniowski:2016trx}
W.~Broniowski, E.~Ruiz~Arriola, K.~Golec-Biernat, {Generalized Valon Model for
  Double Parton Distributions}, Few Body Syst. 57~(6) (2016) 405--410.
\newblock \href {http://arxiv.org/abs/1602.00254} {\path{arXiv:1602.00254}},
  \href {https://doi.org/10.1007/s00601-016-1087-z}
  {\path{doi:10.1007/s00601-016-1087-z}}.

\bibitem{Kasemets:2016nio}
T.~Kasemets, A.~Mukherjee, {Quark-gluon double parton distributions in the
  light-front dressed quark model}, Phys. Rev. D 94~(7) (2016) 074029.
\newblock \href {http://arxiv.org/abs/1606.05686} {\path{arXiv:1606.05686}},
  \href {https://doi.org/10.1103/PhysRevD.94.074029}
  {\path{doi:10.1103/PhysRevD.94.074029}}.

\bibitem{Rinaldi:2016jvu}
M.~Rinaldi, S.~Scopetta, M.~C. Traini, V.~Vento, {Correlations in Double Parton
  Distributions: Perturbative and Non-Perturbative effects}, JHEP 10 (2016)
  063.
\newblock \href {http://arxiv.org/abs/1608.02521} {\path{arXiv:1608.02521}},
  \href {https://doi.org/10.1007/JHEP10(2016)063}
  {\path{doi:10.1007/JHEP10(2016)063}}.

\bibitem{Rinaldi:2016mlk}
M.~Rinaldi, F.~A. Ceccopieri, {Relativistic effects in model calculations of
  double parton distribution function}, Phys. Rev. D 95~(3) (2017) 034040.
\newblock \href {http://arxiv.org/abs/1611.04793} {\path{arXiv:1611.04793}},
  \href {https://doi.org/10.1103/PhysRevD.95.034040}
  {\path{doi:10.1103/PhysRevD.95.034040}}.

\bibitem{Rinaldi:2018zng}
M.~Rinaldi, S.~Scopetta, M.~Traini, V.~Vento, {A model calculation of double
  parton distribution functions of the pion}, Eur. Phys. J. C 78~(9) (2018)
  781.
\newblock \href {http://arxiv.org/abs/1806.10112} {\path{arXiv:1806.10112}},
  \href {https://doi.org/10.1140/epjc/s10052-018-6256-4}
  {\path{doi:10.1140/epjc/s10052-018-6256-4}}.

\bibitem{Courtoy:2019cxq}
A.~Courtoy, S.~Noguera, S.~Scopetta, {Double parton distributions in the pion
  in the Nambu\textendash{}Jona-Lasinio model}, JHEP 12 (2019) 045.
\newblock \href {http://arxiv.org/abs/1909.09530} {\path{arXiv:1909.09530}},
  \href {https://doi.org/10.1007/JHEP12(2019)045}
  {\path{doi:10.1007/JHEP12(2019)045}}.

\bibitem{Broniowski:2019rmu}
W.~Broniowski, E.~Ruiz~Arriola, {Double parton distribution of valence quarks
  in the pion in chiral quark models}, Phys. Rev. D 101~(1) (2020) 014019.
\newblock \href {http://arxiv.org/abs/1910.03707} {\path{arXiv:1910.03707}},
  \href {https://doi.org/10.1103/PhysRevD.101.014019}
  {\path{doi:10.1103/PhysRevD.101.014019}}.

\bibitem{Rinaldi:2020ybv}
M.~Rinaldi, {Double parton correlations in mesons within AdS/QCD soft-wall
  models: a first comparison with lattice data}, Eur. Phys. J. C 80~(7) (2020)
  678.
\newblock \href {http://arxiv.org/abs/2003.09400} {\path{arXiv:2003.09400}},
  \href {https://doi.org/10.1140/epjc/s10052-020-8241-y}
  {\path{doi:10.1140/epjc/s10052-020-8241-y}}.

\bibitem{Courtoy:2020tkd}
A.~Courtoy, S.~Noguera, S.~Scopetta, {Two-current correlations in the pion in
  the Nambu and Jona-Lasinio model}, Eur. Phys. J. C 80~(10) (2020) 909.
\newblock \href {http://arxiv.org/abs/2006.05300} {\path{arXiv:2006.05300}},
  \href {https://doi.org/10.1140/epjc/s10052-020-08470-1}
  {\path{doi:10.1140/epjc/s10052-020-08470-1}}.

\bibitem{Mondal:2019rhs}
C.~Mondal, A.~Mukherjee, S.~Nair, {Double parton distributions for a
  positroniumlike bound state using light-front wave functions}, Phys. Rev. D
  100~(9) (2019) 094002.
\newblock \href {http://arxiv.org/abs/1906.10903} {\path{arXiv:1906.10903}},
  \href {https://doi.org/10.1103/PhysRevD.100.094002}
  {\path{doi:10.1103/PhysRevD.100.094002}}.

\bibitem{CMS:2020cpy}
A.~M. Sirunyan, et~al., {Search for physics beyond the standard model in events
  with jets and two same-sign or at least three charged leptons in
  proton-proton collisions at $\sqrt{s}=$ 13 TeV}, Eur. Phys. J. C 80~(8)
  (2020) 752.
\newblock \href {http://arxiv.org/abs/2001.10086} {\path{arXiv:2001.10086}},
  \href {https://doi.org/10.1140/epjc/s10052-020-8168-3}
  {\path{doi:10.1140/epjc/s10052-020-8168-3}}.

\bibitem{DelFabbro:1999tf}
A.~Del~Fabbro, D.~Treleani, {A Double parton scattering background to Higgs
  boson production at the LHC}, Phys. Rev. D 61 (2000) 077502.
\newblock \href {http://arxiv.org/abs/hep-ph/9911358}
  {\path{arXiv:hep-ph/9911358}}, \href
  {https://doi.org/10.1103/PhysRevD.61.077502}
  {\path{doi:10.1103/PhysRevD.61.077502}}.

\bibitem{Fedkevych:2020cmd}
O.~Fedkevych, A.~Kulesza, {Double parton scattering in four-jet production in
  proton-proton collisions at the LHC}, Phys. Rev. D 104~(5) (2021) 054021.
\newblock \href {http://arxiv.org/abs/2008.08347} {\path{arXiv:2008.08347}},
  \href {https://doi.org/10.1103/PhysRevD.104.054021}
  {\path{doi:10.1103/PhysRevD.104.054021}}.

\bibitem{Ceccopieri:2021luf}
F.~A. Ceccopieri, M.~Rinaldi, {Enlighting the transverse structure of the
  proton via double parton scattering in photon-induced interactions}, Phys.
  Rev. D 105~(1) (2022) L011501.
\newblock \href {http://arxiv.org/abs/2103.13480} {\path{arXiv:2103.13480}},
  \href {https://doi.org/10.1103/PhysRevD.105.L011501}
  {\path{doi:10.1103/PhysRevD.105.L011501}}.

\bibitem{Blok:2022mtv}
B.~Blok, J.~Mehl, {Perturbative Color Correlations in Double Parton Scattering}
  (10 2022).
\newblock \href {http://arxiv.org/abs/2210.13282} {\path{arXiv:2210.13282}}.

\bibitem{Golec-Biernat:2022wkx}
K.~Golec-Biernat, A.~M. Sta\'sto, {Momentum sum rule and factorization of
  double parton distributions}, Phys. Rev. D 107~(5) (2023) 054020.
\newblock \href {http://arxiv.org/abs/2212.02289} {\path{arXiv:2212.02289}},
  \href {https://doi.org/10.1103/PhysRevD.107.054020}
  {\path{doi:10.1103/PhysRevD.107.054020}}.

\bibitem{Diehl:2023jje}
M.~Diehl, F.~Fabry, P.~Ploessl, {Evolution of colour correlated double parton
  distributions: a quantitative study} (10 2023).
\newblock \href {http://arxiv.org/abs/2310.16432} {\path{arXiv:2310.16432}}.

\bibitem{Diehl:2022dia}
M.~Diehl, R.~Nagar, P.~Pl\"o\ss{}l, {Quark mass effects in double parton
  distributions}, JHEP 09 (2023) 100.
\newblock \href {http://arxiv.org/abs/2212.07736} {\path{arXiv:2212.07736}},
  \href {https://doi.org/10.1007/JHEP09(2023)100}
  {\path{doi:10.1007/JHEP09(2023)100}}.

\bibitem{Diehl:2023cth}
M.~Diehl, R.~Nagar, P.~Ploessl, F.~J. Tackmann, {Evolution and interpolation of
  double parton distributions using Chebyshev grids}, Eur. Phys. J. C 83~(6)
  (2023) 536.
\newblock \href {http://arxiv.org/abs/2305.04845} {\path{arXiv:2305.04845}},
  \href {https://doi.org/10.1140/epjc/s10052-023-11692-8}
  {\path{doi:10.1140/epjc/s10052-023-11692-8}}.

\bibitem{Bali:2021gel}
G.~S. Bali, M.~Diehl, B.~Gl\"a\ss{}le, A.~Sch\"afer, C.~Zimmermann, {Double
  parton distributions in the nucleon from lattice QCD}, JHEP 09 (2021) 106.
\newblock \href {http://arxiv.org/abs/2106.03451} {\path{arXiv:2106.03451}},
  \href {https://doi.org/10.1007/JHEP09(2021)106}
  {\path{doi:10.1007/JHEP09(2021)106}}.

\bibitem{Zimmermann:2022emx}
C.~Zimmermann, D.~Reitinger, {Double parton distributions in the nucleon on the
  lattice: Flavor interference effects}, PoS LATTICE2022 (2023) 131.
\newblock \href {http://arxiv.org/abs/2211.14151} {\path{arXiv:2211.14151}},
  \href {https://doi.org/10.22323/1.430.0131} {\path{doi:10.22323/1.430.0131}}.

\bibitem{Zhang:2023wea}
J.-H. Zhang, {Double Parton Distributions from Euclidean Lattice} (4 2023).
\newblock \href {http://arxiv.org/abs/2304.12481} {\path{arXiv:2304.12481}}.

\bibitem{Vary:2009gt}
J.~P. Vary, H.~Honkanen, J.~Li, P.~Maris, S.~J. Brodsky, A.~Harindranath, G.~F.
  de~Teramond, P.~Sternberg, E.~G. Ng, C.~Yang, {Hamiltonian light-front field
  theory in a basis function approach}, Phys. Rev. C 81 (2010) 035205.
\newblock \href {http://arxiv.org/abs/0905.1411} {\path{arXiv:0905.1411}},
  \href {https://doi.org/10.1103/PhysRevC.81.035205}
  {\path{doi:10.1103/PhysRevC.81.035205}}.

\bibitem{Brodsky:1997de}
S.~J. Brodsky, H.-C. Pauli, S.~S. Pinsky, {Quantum chromodynamics and other
  field theories on the light cone}, Phys. Rept. 301 (1998) 299--486.
\newblock \href {http://arxiv.org/abs/hep-ph/9705477}
  {\path{arXiv:hep-ph/9705477}}, \href
  {https://doi.org/10.1016/S0370-1573(97)00089-6}
  {\path{doi:10.1016/S0370-1573(97)00089-6}}.

\bibitem{Honkanen:2010rc}
H.~Honkanen, P.~Maris, J.~P. Vary, S.~J. Brodsky, {Electron in a transverse
  harmonic cavity}, Phys. Rev. Lett. 106 (2011) 061603.
\newblock \href {http://arxiv.org/abs/1008.0068} {\path{arXiv:1008.0068}},
  \href {https://doi.org/10.1103/PhysRevLett.106.061603}
  {\path{doi:10.1103/PhysRevLett.106.061603}}.

\bibitem{Zhao:2014xaa}
X.~Zhao, H.~Honkanen, P.~Maris, J.~P. Vary, S.~J. Brodsky, {Electron g-2 in
  Light-Front Quantization}, Phys. Lett. B 737 (2014) 65--69.
\newblock \href {http://arxiv.org/abs/1402.4195} {\path{arXiv:1402.4195}},
  \href {https://doi.org/10.1016/j.physletb.2014.08.020}
  {\path{doi:10.1016/j.physletb.2014.08.020}}.

\bibitem{Wiecki:2014ola}
P.~Wiecki, Y.~Li, X.~Zhao, P.~Maris, J.~P. Vary, {Basis Light-Front
  Quantization Approach to Positronium}, Phys. Rev. D 91~(10) (2015) 105009.
\newblock \href {http://arxiv.org/abs/1404.6234} {\path{arXiv:1404.6234}},
  \href {https://doi.org/10.1103/PhysRevD.91.105009}
  {\path{doi:10.1103/PhysRevD.91.105009}}.

\bibitem{Chakrabarti:2014cwa}
D.~Chakrabarti, X.~Zhao, H.~Honkanen, R.~Manohar, P.~Maris, J.~P. Vary,
  {Generalized parton distributions in a light-front nonperturbative approach},
  Phys. Rev. D 89~(11) (2014) 116004.
\newblock \href {http://arxiv.org/abs/1403.0704} {\path{arXiv:1403.0704}},
  \href {https://doi.org/10.1103/PhysRevD.89.116004}
  {\path{doi:10.1103/PhysRevD.89.116004}}.

\bibitem{Hu:2020arv}
Z.~Hu, S.~Xu, C.~Mondal, X.~Zhao, J.~P. Vary, {Transverse structure of electron
  in momentum space in basis light-front quantization}, Phys. Rev. D 103~(3)
  (2021) 036005.
\newblock \href {http://arxiv.org/abs/2010.12498} {\path{arXiv:2010.12498}},
  \href {https://doi.org/10.1103/PhysRevD.103.036005}
  {\path{doi:10.1103/PhysRevD.103.036005}}.

\bibitem{Nair:2022evk}
S.~Nair, C.~Mondal, X.~Zhao, A.~Mukherjee, J.~P. Vary, {Basis light-front
  quantization approach to photon}, Phys. Lett. B 827 (2022) 137005.
\newblock \href {http://arxiv.org/abs/2201.12770} {\path{arXiv:2201.12770}},
  \href {https://doi.org/10.1016/j.physletb.2022.137005}
  {\path{doi:10.1016/j.physletb.2022.137005}}.

\bibitem{Nair:2023lir}
S.~Nair, C.~Mondal, X.~Zhao, A.~Mukherjee, J.~P. Vary, {Massless and massive
  photons within basis light-front quantization}, Phys. Rev. D 108~(11) (2023)
  116005.
\newblock \href {http://arxiv.org/abs/2302.13645} {\path{arXiv:2302.13645}},
  \href {https://doi.org/10.1103/PhysRevD.108.116005}
  {\path{doi:10.1103/PhysRevD.108.116005}}.

\bibitem{Jia:2018ary}
S.~Jia, J.~P. Vary, {Basis light front quantization for the charged light
  mesons with color singlet Nambu\textendash{}Jona-Lasinio interactions}, Phys.
  Rev. C 99~(3) (2019) 035206.
\newblock \href {http://arxiv.org/abs/1811.08512} {\path{arXiv:1811.08512}},
  \href {https://doi.org/10.1103/PhysRevC.99.035206}
  {\path{doi:10.1103/PhysRevC.99.035206}}.

\bibitem{Lan:2019vui}
J.~Lan, C.~Mondal, S.~Jia, X.~Zhao, J.~P. Vary, {Parton Distribution Functions
  from a Light Front Hamiltonian and QCD Evolution for Light Mesons}, Phys.
  Rev. Lett. 122~(17) (2019) 172001.
\newblock \href {http://arxiv.org/abs/1901.11430} {\path{arXiv:1901.11430}},
  \href {https://doi.org/10.1103/PhysRevLett.122.172001}
  {\path{doi:10.1103/PhysRevLett.122.172001}}.

\bibitem{Lan:2019rba}
J.~Lan, C.~Mondal, S.~Jia, X.~Zhao, J.~P. Vary, {Pion and kaon parton
  distribution functions from basis light front quantization and QCD
  evolution}, Phys. Rev. D 101~(3) (2020) 034024.
\newblock \href {http://arxiv.org/abs/1907.01509} {\path{arXiv:1907.01509}},
  \href {https://doi.org/10.1103/PhysRevD.101.034024}
  {\path{doi:10.1103/PhysRevD.101.034024}}.

\bibitem{Adhikari:2021jrh}
L.~Adhikari, C.~Mondal, S.~Nair, S.~Xu, S.~Jia, X.~Zhao, J.~P. Vary,
  {Generalized parton distributions and spin structures of light mesons from a
  light-front Hamiltonian approach}, Phys. Rev. D 104~(11) (2021) 114019.
\newblock \href {http://arxiv.org/abs/2110.05048} {\path{arXiv:2110.05048}},
  \href {https://doi.org/10.1103/PhysRevD.104.114019}
  {\path{doi:10.1103/PhysRevD.104.114019}}.

\bibitem{Lan:2021wok}
J.~Lan, K.~Fu, C.~Mondal, X.~Zhao, j.~P. Vary, {Light mesons with one dynamical
  gluon on the light front}, Phys. Lett. B 825 (2022) 136890.
\newblock \href {http://arxiv.org/abs/2106.04954} {\path{arXiv:2106.04954}},
  \href {https://doi.org/10.1016/j.physletb.2022.136890}
  {\path{doi:10.1016/j.physletb.2022.136890}}.

\bibitem{Mondal:2021czk}
C.~Mondal, S.~Nair, S.~Jia, X.~Zhao, J.~P. Vary, {Pion to photon transition
  form factors with basis light-front quantization}, Phys. Rev. D 104~(9)
  (2021) 094034.
\newblock \href {http://arxiv.org/abs/2109.02279} {\path{arXiv:2109.02279}},
  \href {https://doi.org/10.1103/PhysRevD.104.094034}
  {\path{doi:10.1103/PhysRevD.104.094034}}.

\bibitem{Li:2015zda}
Y.~Li, P.~Maris, X.~Zhao, J.~P. Vary, {Heavy Quarkonium in a Holographic
  Basis}, Phys. Lett. B 758 (2016) 118--124.
\newblock \href {http://arxiv.org/abs/1509.07212} {\path{arXiv:1509.07212}},
  \href {https://doi.org/10.1016/j.physletb.2016.04.065}
  {\path{doi:10.1016/j.physletb.2016.04.065}}.

\bibitem{Li:2017mlw}
Y.~Li, P.~Maris, J.~P. Vary, {Quarkonium as a relativistic bound state on the
  light front}, Phys. Rev. D 96 (2017) 016022.
\newblock \href {http://arxiv.org/abs/1704.06968} {\path{arXiv:1704.06968}},
  \href {https://doi.org/10.1103/PhysRevD.96.016022}
  {\path{doi:10.1103/PhysRevD.96.016022}}.

\bibitem{Li:2018uif}
M.~Li, Y.~Li, P.~Maris, J.~P. Vary, {Radiative transitions between $0^{-+}$ and
  $1^{--}$ heavy quarkonia on the light front}, Phys. Rev. D 98~(3) (2018)
  034024.
\newblock \href {http://arxiv.org/abs/1803.11519} {\path{arXiv:1803.11519}},
  \href {https://doi.org/10.1103/PhysRevD.98.034024}
  {\path{doi:10.1103/PhysRevD.98.034024}}.

\bibitem{Lan:2019img}
J.~Lan, C.~Mondal, M.~Li, Y.~Li, S.~Tang, X.~Zhao, J.~P. Vary, {Parton
  Distribution Functions of Heavy Mesons on the Light Front}, Phys. Rev. D
  102~(1) (2020) 014020.
\newblock \href {http://arxiv.org/abs/1911.11676} {\path{arXiv:1911.11676}},
  \href {https://doi.org/10.1103/PhysRevD.102.014020}
  {\path{doi:10.1103/PhysRevD.102.014020}}.

\bibitem{Tang:2018myz}
S.~Tang, Y.~Li, P.~Maris, J.~P. Vary, {$B_c$ mesons and their properties on the
  light front}, Phys. Rev. D 98~(11) (2018) 114038.
\newblock \href {http://arxiv.org/abs/1810.05971} {\path{arXiv:1810.05971}},
  \href {https://doi.org/10.1103/PhysRevD.98.114038}
  {\path{doi:10.1103/PhysRevD.98.114038}}.

\bibitem{Tang:2019gvn}
S.~Tang, Y.~Li, P.~Maris, J.~P. Vary, {Heavy-light mesons on the light front},
  Eur. Phys. J. C 80~(6) (2020) 522.
\newblock \href {http://arxiv.org/abs/1912.02088} {\path{arXiv:1912.02088}},
  \href {https://doi.org/10.1140/epjc/s10052-020-8081-9}
  {\path{doi:10.1140/epjc/s10052-020-8081-9}}.

\bibitem{Mondal:2019jdg}
C.~Mondal, S.~Xu, J.~Lan, X.~Zhao, Y.~Li, D.~Chakrabarti, J.~P. Vary, {Proton
  structure from a light-front Hamiltonian}, Phys. Rev. D 102~(1) (2020)
  016008.
\newblock \href {http://arxiv.org/abs/1911.10913} {\path{arXiv:1911.10913}},
  \href {https://doi.org/10.1103/PhysRevD.102.016008}
  {\path{doi:10.1103/PhysRevD.102.016008}}.

\bibitem{Xu:2021wwj}
S.~Xu, C.~Mondal, J.~Lan, X.~Zhao, Y.~Li, J.~P. Vary, {Nucleon structure from
  basis light-front quantization}, Phys. Rev. D 104~(9) (2021) 094036.
\newblock \href {http://arxiv.org/abs/2108.03909} {\path{arXiv:2108.03909}},
  \href {https://doi.org/10.1103/PhysRevD.104.094036}
  {\path{doi:10.1103/PhysRevD.104.094036}}.

\bibitem{Liu:2022fvl}
Y.~Liu, S.~Xu, C.~Mondal, X.~Zhao, J.~P. Vary, {Angular momentum and
  generalized parton distributions for the proton with basis light-front
  quantization}, Phys. Rev. D 105~(9) (2022) 094018.
\newblock \href {http://arxiv.org/abs/2202.00985} {\path{arXiv:2202.00985}},
  \href {https://doi.org/10.1103/PhysRevD.105.094018}
  {\path{doi:10.1103/PhysRevD.105.094018}}.

\bibitem{Hu:2022ctr}
Z.~Hu, S.~Xu, C.~Mondal, X.~Zhao, J.~P. Vary, {Transverse momentum structure of
  proton within the basis light-front quantization framework}, Phys. Lett. B
  833 (2022) 137360.
\newblock \href {http://arxiv.org/abs/2205.04714} {\path{arXiv:2205.04714}},
  \href {https://doi.org/10.1016/j.physletb.2022.137360}
  {\path{doi:10.1016/j.physletb.2022.137360}}.

\bibitem{Peng:2022lte}
T.~Peng, Z.~Zhu, S.~Xu, X.~Liu, C.~Mondal, X.~Zhao, J.~P. Vary, {Basis
  light-front quantization approach to \ensuremath{\Lambda} and
  \ensuremath{\Lambda}c and their isospin triplet baryons}, Phys. Rev. D
  106~(11) (2022) 114040.
\newblock \href {http://arxiv.org/abs/2208.00355} {\path{arXiv:2208.00355}},
  \href {https://doi.org/10.1103/PhysRevD.106.114040}
  {\path{doi:10.1103/PhysRevD.106.114040}}.

\bibitem{Kaur:2023lun}
S.~Kaur, S.~Xu, C.~Mondal, X.~Zhao, J.~P. Vary, {Spatial imaging of proton via
  leading-twist nonskewed GPDs with basis light-front quantization}, Phys. Rev.
  D 109~(1) (2024) 014015.
\newblock \href {http://arxiv.org/abs/2307.09869} {\path{arXiv:2307.09869}},
  \href {https://doi.org/10.1103/PhysRevD.109.014015}
  {\path{doi:10.1103/PhysRevD.109.014015}}.

\bibitem{Zhu:2023nhl}
Z.~Zhu, T.~Peng, Z.~Hu, S.~Xu, C.~Mondal, X.~Zhao, J.~P. Vary, {Transverse
  momentum structure of strange and charmed baryons: A light-front Hamiltonian
  approach}, Phys. Rev. D 108~(3) (2023) 036009.
\newblock \href {http://arxiv.org/abs/2304.05058} {\path{arXiv:2304.05058}},
  \href {https://doi.org/10.1103/PhysRevD.108.036009}
  {\path{doi:10.1103/PhysRevD.108.036009}}.

\bibitem{Zhang:2023xfe}
Z.~Zhang, Z.~Hu, S.~Xu, C.~Mondal, X.~Zhao, J.~P. Vary, {Twist-3 generalized
  parton distribution for the proton from basis light-front quantization},
  Phys. Rev. D 109~(3) (2024) 034031.
\newblock \href {http://arxiv.org/abs/2312.00667} {\path{arXiv:2312.00667}},
  \href {https://doi.org/10.1103/PhysRevD.109.034031}
  {\path{doi:10.1103/PhysRevD.109.034031}}.

\bibitem{Liu:2024umn}
Y.~Liu, S.~Xu, C.~Mondal, X.~Zhao, J.~P. Vary, {Skewed generalized parton
  distributions of proton from basis light-front quantization} (3 2024).
\newblock \href {http://arxiv.org/abs/2403.05922} {\path{arXiv:2403.05922}}.

\bibitem{Kaur:2024iwn}
S.~Kaur, J.~Wu, Z.~Hu, J.~Lan, C.~Mondal, X.~Zhao, J.~P. Vary, {Quark and gluon
  distributions in \ensuremath{\rho}-meson from basis light-front
  quantization}, Phys. Lett. B 851 (2024) 138563.
\newblock \href {http://arxiv.org/abs/2401.03480} {\path{arXiv:2401.03480}},
  \href {https://doi.org/10.1016/j.physletb.2024.138563}
  {\path{doi:10.1016/j.physletb.2024.138563}}.

\bibitem{Yu:2024mxo}
H.~Yu, Z.~Hu, S.~Xu, C.~Mondal, X.~Zhao, J.~P. Vary,
  {Transverse-momentum-dependent gluon distributions of proton within basis
  light-front quantization} (3 2024).
\newblock \href {http://arxiv.org/abs/2403.06125} {\path{arXiv:2403.06125}}.

\bibitem{Xu:2023nqv}
S.~Xu, C.~Mondal, X.~Zhao, Y.~Li, J.~P. Vary, {Quark and gluon spin and orbital
  angular momentum in the proton}, Phys. Rev. D 108~(9) (2023) 094002.
\newblock \href {https://doi.org/10.1103/PhysRevD.108.094002}
  {\path{doi:10.1103/PhysRevD.108.094002}}.

\bibitem{Lin:2023ezw}
B.~Lin, S.~Nair, S.~Xu, Z.~Hu, C.~Mondal, X.~Zhao, J.~P. Vary, {Generalized
  parton distributions of gluon in proton: A light-front quantization
  approach}, Phys. Lett. B 847 (2023) 138305.
\newblock \href {http://arxiv.org/abs/2308.08275} {\path{arXiv:2308.08275}},
  \href {https://doi.org/10.1016/j.physletb.2023.138305}
  {\path{doi:10.1016/j.physletb.2023.138305}}.

\bibitem{Zhu:2023lst}
Z.~Zhu, Z.~Hu, J.~Lan, C.~Mondal, X.~Zhao, J.~P. Vary, {Transverse structure of
  the pion beyond leading twist with basis light-front quantization}, Phys.
  Lett. B 839 (2023) 137808.
\newblock \href {http://arxiv.org/abs/2301.12994} {\path{arXiv:2301.12994}},
  \href {https://doi.org/10.1016/j.physletb.2023.137808}
  {\path{doi:10.1016/j.physletb.2023.137808}}.

\bibitem{Harindranath:1996hq}
A.~Harindranath, {An Introduction to light front dynamics for pedestrians}, in:
  {International School on Light-Front Quantization and Non-Perturbative QCD
  (To be followed by the Workshop 3-14 Jun 1996)}, 1996.
\newblock \href {http://arxiv.org/abs/hep-ph/9612244}
  {\path{arXiv:hep-ph/9612244}}.

\bibitem{Zhao:2013cma}
X.~Zhao, A.~Ilderton, P.~Maris, J.~P. Vary, {Scattering in Time-Dependent Basis
  Light-Front Quantization}, Phys. Rev. D 88 (2013) 065014.
\newblock \href {http://arxiv.org/abs/1303.3273} {\path{arXiv:1303.3273}},
  \href {https://doi.org/10.1103/PhysRevD.88.065014}
  {\path{doi:10.1103/PhysRevD.88.065014}}.

\bibitem{Brodsky:2014yha}
S.~J. Brodsky, G.~F. de~Teramond, H.~G. Dosch, J.~Erlich, {Light-Front
  Holographic QCD and Emerging Confinement}, Phys. Rept. 584 (2015) 1--105.
\newblock \href {http://arxiv.org/abs/1407.8131} {\path{arXiv:1407.8131}},
  \href {https://doi.org/10.1016/j.physrep.2015.05.001}
  {\path{doi:10.1016/j.physrep.2015.05.001}}.

\end{thebibliography}

\end{document}